\documentclass[preprint,11pt,numbers,sort&compress]{elsarticle}
\usepackage{subcaption}
\usepackage{amsmath,amssymb,amsthm,amsfonts}
\counterwithout{equation}{section}
\usepackage{mathabx}
\usepackage{mathrsfs}
\usepackage{bm}
\usepackage{graphicx}
\usepackage{color,xcolor}
\usepackage{tabularx}
\usepackage{tikz}
\usepackage{pgflibraryshapes}
\usepackage{graphicx,psfrag}
\usepackage[colorlinks,linkcolor=blue]{hyperref}
\usepackage{enumitem}
\usepackage{changes}
\makeatletter
\@namedef{Changes@AuthorColor}{blue}
\colorlet{Changes@Color}{blue}
\makeatother

\makeatletter
\hypersetup{colorlinks=true}
\AtBeginDocument{\@ifpackageloaded{hyperref}
	{\def\@linkcolor{magenta}
		\def\@anchorcolor{black}
		\def\@citecolor{teal}
		\def\@filecolor{cyan}
		\def\@urlcolor{magenta}
		\def\@menucolor{red}
		\def\@pagecolor{cyan}
		\begingroup
		\@makeother\`%
		\@makeother\=%
		\edef\x{%
			\edef\noexpand\x{%
				\endgroup
				\noexpand\toks@{%
					\catcode 96=\noexpand\the\catcode`\noexpand\`\relax
					\catcode 61=\noexpand\the\catcode`\noexpand\=\relax
				}%
			}%
			\noexpand\x
		}%
		\x
		\@makeother\`
		\@makeother\=
	}{}}
\makeatother

\makeatletter \oddsidemargin 0.0cm \evensidemargin 0.0cm
\newcommand{\be}{\begin{equation}}
\newcommand{\en}{\end{equation}}

\def\bm#1{\mbox{\boldmath{$#1$}}}

\theoremstyle{plain}

\newtheorem{theorem*}{Theorem}

\theoremstyle{definition}

\begin{document}

\begin{frontmatter}
		
\title{{\bf Surface elasticity effect on Plateau--Rayleigh instability in soft solids}}
		
\author[mymainaddress]{Pingping Zhu}
\author[mymainaddress]{Dun Li}
\author[mysecondaryaddress]{Xiang Yu\corref{mycorrespondingauthor}}
\ead{yuxiang@dgut.edu.cn}
\author[mymainaddress]{Zheng Zhong\corref{mycorrespondingauthor}}
\cortext[mycorrespondingauthor]{Corresponding author }
\ead{zhongzheng@hit.edu.cn}
		
\address[mymainaddress]{School of Science, Harbin Institute of Technology, Shenzhen 518055, PR China}
		
\address[mysecondaryaddress]{Department of Mathematics, School of Computer Science and Technology, Dongguan University of Technology, Dongguan, 523808, PR China}
		
\begin{abstract}
Soft solids exhibit instability and develop surface undulations due to surface effects, a phenomenon known as the elastic Plateau-Rayleigh (PR) instability, driven by the interplay of surface and bulk elasticity.  Previous studies on the PR instability in solids mainly focused on the case of constant surface tension and ignored the effect of surface elasticity. It has been shown by experiments that the surface effects in solid-like materials depend both on the surface tension and surface elasticity, but little is known about the role of the latter in the elasto-capillary instabilities in soft solids. Here, we conduct an in-depth exploration of the effect of surface elasticity on the PR instability in an elastic cylinder by coupling theoretical and numerical methods. We derive an asymptotically consistent one-dimensional (1d) model to characterize the PR instability from three-dimensional (3d) nonlinear bulk-surface elasticity, and develop a new finite-element (FE) scheme for simulating  3d deformations of the bulk-surface system. The initiation and evolution of the PR instability are obtained analytically with the aid of the 1d model. The 1d results are further validated by the 3d FE simulations. By synthesizing the 1d analytic solutions and 3d numerical results, the effects of surface elasticity, surface compressibility, surface tension, axial force and geometrical size on the PR instability are thoroughly elucidated. It is found that a given surface parameter can respectively induce distinct effects on the threshold and the amplitude of the PR instability. For instance, when the axial force is controlled to vary, we find that strong surface elasticity and surface incompressibility hinder the initiation of necking but advance its growth, while the surface tension plays an opposite role in this process. Our results can be applied to calibrate surface parameters for solid-like materials and develop constitutive models for elastic surfaces.
\end{abstract}
		
\end{frontmatter}

\section{Introduction}

In the past few decades, soft materials, structural miniaturization and micro/nano-technology have been continuously flourishing \cite{C2022}, highlighting the significant importance of surface effects in modeling mechanical behaviors of soft solids, especially in micro and nano scales. Surface effects arise as a result of the distinct elastic properties of the surface and bulk solid. The competition between surface energy and bulk elastic energy can induce numerous elasto-capillary instabilities in solids. For instance, soft elastic cylinders can experience instabilities and develop surface undulations (e.g., necking, bulging and beading) under surface effects, akin to the Plateau--Rayleigh instabilities in liquids \cite{Li2024a,G2024a}. Other elasto-capillary instabilities include creasing, wrinkling, cavitation, and so on \cite{K2013,R2023,Mora2019,Liu2019,E2023,van2021,T2025}. Materials implicated in these instabilities have been found to include soft gels \citep{MT1992,Mora2010,Mora2011b,Z2007}, nerve fibers \citep{Markin1999}, electrospinning nanofibers \citep{Fong1999},  high entangled polymer nanofibers \cite{sattler2008},  semiconductor fibres during thermal drawing \citep{Gao2024}, microfluidic devices during fabrication \citep{K2020}, etc. Surface effects have also been observed to affect the functions of natural biological systems \citep{C1997,B1997} and micro-electro-mechanical systems \citep{Hui2002}, and change morphologies of ultrasoft gels \citep{Mora2013,Mora2015,Zhu2023} and nano-scale structures \citep{Wang2021}. Therefore, understanding the mechanisms behind these complex surface-effect-induced phenomena is crucial, both for preventing resulting instabilities and utilizing surface effects to realize specific functions in  applications such as biomedical engineering, nanomaterials, soft robotics, adhesives, etc. 
	
After the beading patterns was first spotted in shrinking gels by Matsuo and Tanaka \cite{MT1992}, Kuroki and Sekimoto \cite{K1994} connected the beading formation with the deswollen coating at the gel surface and interpreted the effect of coating as surface tension. After that, several theoretical works \citep{B1996,B2003} investigated the PR instability in soft elastic cylinders in the scope of linear elastic theory. It has become a consensus that the elasto-capillary number, defined by $\gamma/(\mu A)$ with $\gamma$, $\mu$ and $A$ being the surface tension, bulk shear modulus and radius of the cylinder respectively, controls the onset of instability \cite{Bico2018}.

As a series of elasto-capillary phenomena in soft solids were reported by Mora and his coworkers \citep{Mora2010,Mora2011b,Mora2013,Mora2015}, a surge of interest in revealing  the role of the surface tension $\gamma$ in the elasto-capillary instabilities  has transpired in these years. The approaches include numerical simulations \citep{HB2014,Feng2022,Xu2023,Wang2024}, bifurcation analysis \citep{TC2015,XB2017,Fu2021,Emery2021,Emery2021b,Emery2021c,DJ2022,Emery2023,TB2024} and dimension reduction models \citep{LA2020a}.  These works are based on the assumption that the surface energy density is a constant, which leads to a constant, isotropic surface stress (namely, the surface tension). This assumption is valid for fluids, yet it is not appropriate for solids. In a fluid, the atoms or molecules possess extremely high mobility such that the surface area can be reduced without stretching bonds between particles \citep{HB2014}. It is contrary for most solids, in which the atoms or molecules can not move freely. Stretching the bonds between particles on the surface is necessary for solids to change their surface areas. As a result, the surface energy density or the surface stress of a solid is non-constant and non-isotropic but depends on the deformation of the surface \cite{S1950,Muller2004}. This has been indicated by the fact that under the assumption of constant surface tension, the experimentally identified values of surface tension in the same solid material setting are distributed in a rather wide range \citep{ZB2021}. An example can be found in silicone in which the surface tension is reported to range from $\sim20$ to $100$ mN/m.
	
Direct evidence for strain-dependent surface stress is first presented by Xu et al. \cite{Xu2017} through a static wetting experiment \cite{A2020}. Bain et al. \cite{Bain2021} alternatively spotted the strain dependence of surface effect by macroscopic experiment, in which the surface topography of silicone gels is quantified as a function of applied stretch. Via these experiments, the surface stress was found to depend on the applied stretch, characterized by two surface parameters: surface tension $\gamma$ and surface shear modulus $\mu_\text{s}$. The former refers to the residual surface stress when  no stretch is applied and the latter describe how the surface stress is changed by  the surface strain. Another surface parameter affecting the surface properties comes out to be surface compressibility. It is quantified by the surface Poisson's ratio $\nu_\text{s}$, which can naturally lie in a wide range (e.g. lipid bilayers \citep{S2019}) or be adjusted by adding specific coatings to the surface of the bulk.  Zafar and Nasu \cite{ZB2021} proposed a direct method to extract the set of surface parameters $(\gamma, \mu_\text{s}, \nu_\text{s})$ from direct tests of a soft cylinder. Heyden et al. \cite{Heyden2022} provided a robust method for calibrating the three surface parameters from deformations of microscopic liquid droplets embedded in soft solids. They determined the surface parameters from the experiment results from Style et al. \cite{Style2015} as $\gamma=3.9$mN/m, $\mu_\text{s}=2$mN/m, $\nu_\text{s}=0.99$, which yields much better predictions than that by using the setting of constant surface stress ($\gamma$, $0$, $0$). Conducting experiments and accurately quantifying the strain-dependent surface stress is challenging and ongoing \cite{Jensen2017,Z2021}.

As the assumption of constant surface stress is not very suitable for many soft solids, investigating the elasto-capillary behaviors of solids due to strain-dependent surface stresses is imperative. It is essential to determine the roles of surface elasticity and surface compressibility in bulk-surface systems. Bakiler et al. \cite{BD2023} have taken the first step, towards formulating a three-dimensional (3d) theoretical framework for axisymmetric deformation of a hyperelastic cylinder with a generic surface energy. They conducted a linear analysis to derive the bifurcation condition and found that the surface elasticity and surface incompressibility play opposite roles compared with the surface tension on the onset of the PR instability in soft cylinders. However, due to complex coupling of bulk-surface elasticity, the post-bifurcation behavior remains unexplored either numerically or analytically in their work.

In this paper, we perform an in-depth exploration into the effects of surface elasticity and surface compressibility on the PR instability in a hyperelastic cylinder subjected to surface stress and axial stretching. Our analysis extends beyond the linear bifurcation regime studied by Bakiler et al. \cite{BD2023}  and is applicable to the entire post-bifurcation regime. Specifically, in the framework of finite surface elasticity theory initiated by Gurtin and Ian Murdoch \cite{gurtin1975continuum} and developed by many researchers \citep{duan2005size,huang2006theory,steinmann2008boundary,javili2009finite,javili2010finite,papastavrou2013mechanics}, we carry out the dimension reduction using the variational asymptotic method \citep{B1979,A2016,LA2020b,YC2024}, and develop an effective finite-element (FE) scheme based on the Rayleigh-Ritz method \citep{RR2014}.   The reduction method used is a combination of variational principles and asymptotic approaches, and has been applied to  derive highly-accurate 1d models for various nonlinear slender structures  \citep{LA2018,LA2020b,LA2020a,A2021,KA2023,yu2023,yu2025}. For the FE scheme, we point out that in numerical simulations it is not an easy task to impose the boundary conditions at the two ends of the cylinder, since the surface stress is concentrated at the edges of the end surfaces, resulting in a non-uniform, singular stress distribution at the two ends. Our work contributes in the following aspects:
\begin{enumerate}[label=(\roman*)]
\item We derive a 1d reduced model from the 3d nonlinear model that incorporates the surface elasticity effects. The 1d model can capture the PR instability in the cylinder in the entire post-bifurcation regime with remarkable accuracy, due to its asymptotic consistency.
\item We present a new approach to enforce the end conditions and propose an efficient and easy-to-use FE scheme for simulating  3d deformations of the bulk-surface system. One significant advantage of the numerical scheme is its capability to follow the bifurcation process without imposing an initial imperfection, a triggering condition typically required in FE software like Abaqus.
\item  We elaborate on the effects of surface shear modulus, surface compressibility, surface tension, axial force and geometrical size  on the PR instability. These new findings are useful in the measurement of surface parameters and constitutive modeling of elastic surfaces \cite{Gurtin1975,S1999,Duan2009,Hui2020}.
\end{enumerate}

The structure of this paper is as follows. In Section \ref{sec:3d}, we describe the 3d nonlinear elasticity model for a hyperelastic cylinder subjected to surface stress. The solution for the homogeneous deformation of the cylinder is summarized in Section \ref{sec:hom}. Section \ref{sec:1d} is devoted to the derivation of the 1d gradient model by asymptotic reduction. Section \ref{sec:finite-element method} develops the FE scheme of the 3d formulation. The asymptotically analytical and FE results are presented in Section \ref{sec:analysis}, accompanied by a thorough discussion. Finally, a conclusion is drawn in Section \ref{sec:con}.

\section{Three-dimensional finite-strain model incorporating surface elasticity}\label{sec:3d}

We start with establishing the 3d nonlinear bulk-surface elasticity. The  total potential energy functional for the 3d deformation of a solid cylinder subjected to axial force and strain-dependent surface stress is formulated.
	
\begin{figure}[h!]
\centering
\includegraphics[width=0.9\linewidth]{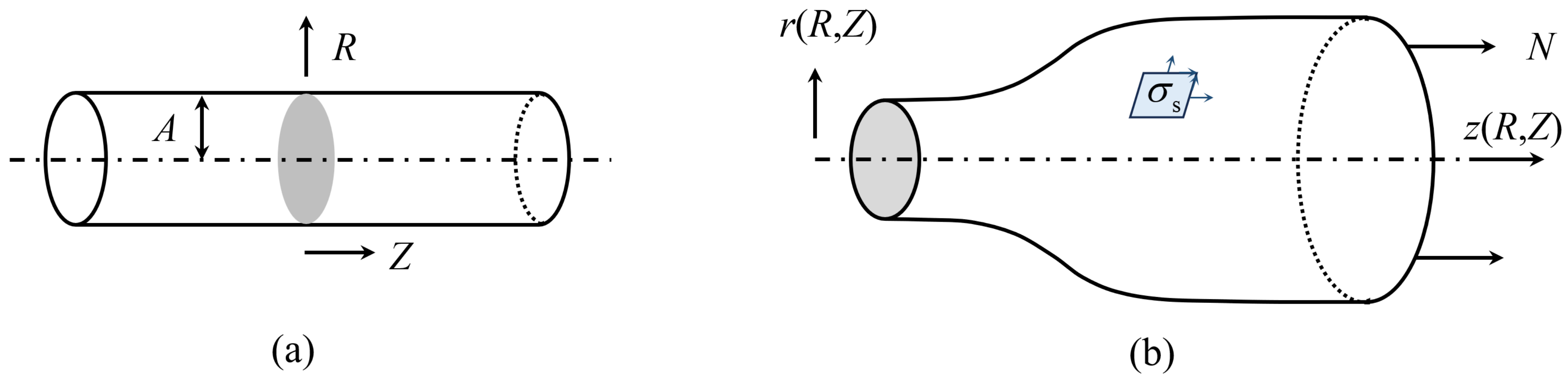}
\caption{A hyperelastic cylinder in (a) reference (undeformed) configuration and (b) current configuration. The cylinder is subjected to an axial force and surface stress.}
\label{fig:cylinder}
\end{figure}

Consider a  hyperelastic solid  cylinder that has a  length $2L$ and radius $A$ in its reference configuration; see Fig. \ref{fig:cylinder}(a). The aspect ratio $\varepsilon=A/(2L)$ is assumed to be small, thus $\varepsilon\ll 1$. The cylinder deforms axisymmetrically due to the combined action of an axial force applied at its two ends and surface stress on its outer surface, as shown in Fig. \ref{fig:cylinder}(b). Here the surface stress  is derived from the constitutive law of the surface, which is equipped its own free energy.
	
Cylindrical coordinates $(R,\Theta,Z)$ and $(r,\theta,z)$ are employed for the reference and current configurations, respectively, and the common set of standard basis vectors for these two sets of cylindrical coordinates are denoted by $\{\bm{e}_r,\bm{e}_\theta,\bm{e}_z\}$. The position of a material point in the reference configuration is given by
\begin{align}
\bm{X}=Z\bm{e}_z+R\bm{e}_r.
\end{align}
After deformation,  the material point $\bm{X}$ moves to a new position
\begin{align}
\bm{x}=z(Z,R)\bm{e}_z+r(Z,R)\bm{e}_r.
\end{align}
The deformation gradient $\bm{F}=\partial\bm{x}/\partial\bm{X}$ is then calculated by as
\begin{align}\label{eq:F}
\bm{F}=\begin{pmatrix}
	\frac{r}{R} & 0 & 0\\
	0 & z_Z & z_R\\
	0 & r_Z & r_R
\end{pmatrix},
\end{align}
where $z_Z=\partial z/\partial Z$, $z_R=\partial z/\partial R$, etc., and here and henceforth  the matrix representations of tensors are always written with respect to the ordered basis $(\bm{e}_\theta,\bm{e}_z,\bm{e}_r)$.

Describing the deformation of the surface without involving that of the bulk solid is done in the framework of surface elasticity \cite{Gurtin1975,S1999,Duan2009,Hui2020}. The position function $\bm{Y}$ on the outer  surface in the undeformed state is identified with the restriction of $\bm{X}$ to $R=A$ as
		\begin{align}
		\bm{Y}=\bm{X}|_{R=A}=Z\bm{e}_z+A\bm{e}_r.
		\end{align}
		The outer surface is convected by the deformation of the bulk cylinder, such that the reference position on the surface  $\bm{Y}$ is mapped to 
		\begin{align}
		\bm{y}=\bm{x}|_{R=A}=z(Z,A)\bm{e}_z+r(Z,A)\bm{e}_r,
		\end{align}
under the deformation. 
		
The {\it surface deformation gradient} is then defined by  
		\begin{align}
		\bm{F}_\text{s}=\frac{\partial\bm{y}}{\partial\bm{Y}},
		\end{align}
		where the letter ``s"  signifies association with the surface \cite{S1999}. 
		Using this definition, we have 
		\begin{align}\label{eq:Fs}
		\bm{F}_\text{s}=\begin{pmatrix}
		\frac{r(Z,A)}{A} & 0 & 0\\
		0 & z_Z(Z,A) & 0\\
		0 & r_Z(Z,A) & 0
		\end{pmatrix}.
		\end{align}
		
The structure of $\bm{F}_\text{s}$ explicitly shows its rank-2 nature. The zero column corresponds to the radial direction $\bm{e}_r$, which is projected out by  $\bm{1}=\bm{I}-\bm{e}_r\otimes\bm{e}_r$ onto the tangent plane of the undeformed outer surface  with $\bm{I}$ being the 3d identity tensor.  
One can see from the matrix forms \eqref{eq:F} and \eqref{eq:Fs} that the bulk deformation gradient $\bm{F}$ and the surface deformation gradient $\bm{F}_\text{s}$ are related by
		\begin{align}\label{eq:FsF}
		\bm{F}_\text{s}=\bm{F}|_{R=A}\bm{1}.
		\end{align}
While $\bm{F}$ is a full-rank $3 \times 3$ tensor, $\bm{F}_\text{s}$ is inherently rank-deficient due to the projection tensor $\bm{1}$. This rank reduction (from $3$ to $2$) signifies that $\bm{F}_\text{s}$ operates purely within the 2d tangent space of the surface, whereas $\bm{F}$ captures 3d bulk changes.

The surface elasticity framework requires separate constitutive laws for the bulk and surface. Assume that the constitutive behavior of the bulk solid is described by a strain energy function $W(\lambda_1,\lambda_2,\lambda_3)$, where $\lambda_i$, $i=1,2,3$ denote the three principal stretches (eigenvalues of $\sqrt{\bm{F}^T\bm{F}}$) and we identify the indices $1, 2, 3$ such that in homogeneous deformations they coincide with the $\theta$-, $z$- and $r$-directions, respectively. As mentioned earlier, the outer surface is assumed to have its own constitutive law, characterized by a surface energy function $\varGamma(\lambda^\text{s}_1,\lambda^\text{s}_2)$, where $\lambda^\text{s}_1$ and $\lambda^\text{s}_2$ signify the two principal stretches of the surface (eigenvalues of $\sqrt{\bm{F}_\text{s}^T\bm{F}_\text{s}}$). For instance,  in the case of liquid-like surface tension, we have $\varGamma(\lambda^\text{s}_1,\lambda^\text{s}_2)=\gamma\lambda_1^\text{s}\lambda_2^\text{s}$ with $\gamma$ being  a constant.  The total potential energy of the cylinder is composed of the bulk energy, surface energy and load potential, which reads, after scaling by $2\pi$,
	\begin{align}\label{eq:E}
	\mathcal{E}[z,r]=\int_{-L}^{L}\Big[\int_0^A \Big(W(\lambda_1,\lambda_2,\lambda_3)-N z_Z\Big)R\,dR+A\varGamma(\lambda^\text{s}_1,\lambda^\text{s}_2) \Big]\,dZ,
	\end{align}
	where $N$ is the resultant axial force per unit reference cross-sectional area. The term $\int_0^A (W(\lambda_1,\lambda_2,\lambda_3)-N z_Z)2\pi R\,dR$  integrates the strain energy density $W$ (due to deformation) and the work done by axial load $N$ over the cross-sectional area. The term  $\int_{-L}^{L}2\pi A\varGamma(\lambda^\text{s}_1,\lambda^\text{s}_2)\,dZ$ represents the total surface energy of the cylinder. This energy functional \eqref{eq:E} corresponds to the 3d nonlinear model that incorporates surface elasticity, which will be used as the starting point of the subsequent dimension reduction.

\section{Homogeneous deformation} \label{sec:hom}
	
Let us first analyze homogeneous deformations for which the stretches are constants. Such a deformation takes the form
	\begin{align}\label{eq:hom}
	z_\text{hom}(Z,R)=\lambda Z,\quad r_\text{hom}(Z,R)=a R,
	\end{align}
where $\lambda$ and $a$ are the constant axial stretch and transverse stretch, respectively. The corresponding deformation gradient follows as
	\begin{align}
	\bm{F}=a\bm{e}_\theta\otimes\bm{e}_\theta+\lambda\bm{e}_z\otimes\bm{e}_z+a\bm{e}_r\otimes\bm{e}_r,
	\end{align}
and consequently, the three principal stretches are
	\begin{align}\label{eq:ps}
	\lambda_1=\lambda_3=a,\quad \lambda_2=\lambda.
	\end{align}
In view of \eqref{eq:FsF}, the surface deformation gradient is given by
	\begin{align}
	\bm{F}_\text{s}=a\bm{e}_\theta\otimes\bm{e}_\theta+\lambda\bm{e}_z\otimes\bm{e}_z,
	\end{align}
and the two  principal stretches of the surface are
	\begin{align}\label{eq:pss}
	\lambda_1^\text{s}=a=\lambda_1,\quad \lambda_2^\text{s}=\lambda=\lambda_2.
	\end{align}

Incorporating \eqref{eq:ps} and \eqref{eq:pss} into \eqref{eq:E}, we see that the total potential energy of the homogeneous deformation per unit reference length is given by
	\begin{align}\label{eq:Ehom}
	\varPhi(a,\lambda)=\frac{1}{2}A^2w(a,\lambda)+A \varGamma(a,\lambda)-\frac{1}{2}A^2N \lambda,
	\end{align}
	where $w(a,\lambda):=W(a,\lambda,a)$ is a reduced strain energy function. Equilibrium is satisfied by requiring $\partial \varPhi/\partial a=0$ and $\partial \varPhi/\partial \lambda=0$, which lead to
	\begin{align}
	&w_1(a,\lambda)+\frac{2}{A}\varGamma_1(a,\lambda)=0, \label{eq:L1}\\
	&N= w_2(a,\lambda)+\frac{2}{A}\varGamma_2(a,\lambda), \label{eq:L2}
	\end{align}
	where the subscripts indicate partial derivatives, i.e., $w_i=\partial w/\partial\lambda_i$ and $\varGamma_i=\partial \varGamma/\lambda^\text{s}_i$, $i=1,2$. Eqs. \eqref{eq:L1} and \eqref{eq:L2} determine the homogeneous deformation completely. Once the axial force $N$ and the strain energy functions of the bulk  and surface are specified, the deformation parameters $a$ and $\lambda$ can be identified by solving these two equations.
	
\section{Asymptotically consistent one-dimensional model}\label{sec:1d}
	
Due to surface stress and axial stretching, the homogeneous deformation may become unstable and a localized necking or bulging instability takes place when the loading parameter reaches a critical value. In the sequel, we derive a 1d reduced model to characterize the necking and bulging deformations using the variational asymptotic method. The 1d model is an ordinary differential equation (ODE) involving only the function of axial stretch.

\subsection{Expansion of the energy functional}
Considering that the necking/bulging deformation varies slowly in axial direction due to its long wavelength, it is natural to define a ``far distance'' variable $S$ through
	\begin{align}
	S=\varepsilon Z,
	\end{align}
	where we recall that $\varepsilon=A/(2L)$ is the small aspect ratio. The slowly varying feature of the necking/bulging deformation implies that the dependence of state variables on $Z$ is through $S$. Guided by the scalings in previous works  \citep{A2016,LA2020a,yu2023}, we look for an asymptotic solution of the form
	\begin{align}\label{eq:asysol}
	\begin{split}
	&z(Z,R)=\frac{1}{\varepsilon}\int_0^S \lambda(\tilde{S})d\tilde{S}+\varepsilon v^*(S,R)+O(\varepsilon^3),\\
	&r(Z,R)=a(S)R+\varepsilon^2 u^*(S,R)+O(\varepsilon^4),
	\end{split}
	\end{align}
	where $v^*$ satisfies the constraint 
	\begin{align}\label{eq:vcon}
	\int_0^A v^*(S,R)R\,dR=0,
	\end{align}
	so that $\lambda$ represents the average of the axial stretch over the cross-section.  In this section, by a slight abuse of notation, we use the same notation to represent quantities of the inhomogeneous solution. For instance, $\lambda$ and $a$ now stand for $\lambda(S)$ and $a(S)$, respectively.

Based on the expansion \eqref{eq:asysol}, the deformation gradients and the principal stretches for both the bulk and the outer surface are truncated to $O(\varepsilon^2)$ (given in \ref{app:trunction}). Then, by substituting the truncated principal stretches \eqref{eq:lambda} and \eqref{eq:lambdas} into the 3d potential energy  \eqref{eq:E} and expanding the resulting expression to order $\varepsilon^2$ with the use of \eqref{eq:vcon}, we see that the energy potential $\mathcal{E}$ can be written in terms of the unscaled variables as
	\begin{align}\label{eq:EE}
	\mathcal{E}[\lambda,a,u,v]=\int_{-L}^L \varPhi(a(Z),\lambda(Z))\,dZ+\mathcal{E}_2+O(L\varepsilon^3),
	\end{align}
	where $\mathcal{E}_2$ represents terms of order $L\varepsilon^2$ and is given by
	\begin{align}\label{eq:EE2}
	\begin{split}
	\mathcal{E}_2=&\int_{-L}^L\Big[ \int_0^A \Big(w_2 \frac{\lambda (v^{2}_R+R^2 a'(Z)^2 )+2 R a a'(Z) v_R}{2(\lambda^2-a^2)}-\frac{w_1}{2} \frac{a (v^{2}_R+R^2 a'(Z)^2 )+2 R \lambda a'(Z) v_R}{2(\lambda^2-a^2)}\Big)R\,dR\\
	&+\int_0^A \frac{w_1}{2} (u+R u_R)\,dR+\varGamma_1 u(Z,A)+ \varGamma_2 A v_Z(Z,A)+\varGamma_2  \frac{A^3 a'(Z)^2}{2\lambda}
	\Big]\,dZ.
	\end{split}
	\end{align}
In the above expressions, $v(Z,R)=\varepsilon v^*(S,R)$ and $u(Z,R)=\varepsilon^2 u^*(S,R)$ denote the unscaled displacements, the quantities $w_i$ and $\varGamma_i$, $i=1,2$ are evaluated at $(a(Z),\lambda(Z))$, such that $w_1(a,\lambda)/2=W_1(a,\lambda,a)=W_3(a,\lambda,a)$ and $w_2(a,\lambda)=W_2(a,\lambda,a)$, and here and hereafter we write $a(Z)$ for $a(S)$ and $\lambda(Z)$ for $\lambda(S)$.  The purpose of introducing $S$ above is to identify all terms of order $\varepsilon^2$ that should be kept in \eqref{eq:EE2}. It will no longer appear in the subsequent analysis.
	
\subsection{Optimal correction}
	
The truncated energy functional \eqref{eq:EE} defines a variational problem for the state variables $\lambda$, $a$, $u$ and $v$, and we need to find the optimal solutions such that \eqref{eq:EE} is stationary. A stepwise variational method that stems from the variational asymptotic method \citep{B1979,A2016,LA2020b,YC2024} is applied to do this, which simplifies the process of finding stationary points of  the variational problem.

First, for a given axial-stretch function $\lambda(Z)$, the leading-order term of the energy functional \eqref{eq:EE} is minimized when $\partial \varPhi/\partial a=0$, which yields the transverse equilibrium \eqref{eq:L1}. Given that the strong ellipticity condition is satisfied pointwise, this equation defines an implicit function for the transverse stretch $a=a(\lambda)$ by the implicit function theorem, which means that $a(\lambda)$ is the solution to
	\begin{align}\label{eq:implicit}
	w_1(a(\lambda),\lambda)+\frac{2}{A}\varGamma_1(a(\lambda),\lambda)=0.
	\end{align}
An important feature of the stepwise minimization is that it is still asymptotically consistent to replace $a(Z)$ by its leading-order approximation $a(\lambda(Z))$ in \eqref{eq:EE} \citep{B1979,YC2024}. Thus we shall use $a=a(\lambda)$ in the sequel.
	
The next step is to minimize $\mathcal{E}_2$ with respect to the functions $u$ and $v$. Note that $w_1$ is independent of $R$, so the terms related to $u$ in \eqref{eq:EE2} can be simplified to
	\begin{align}
	\begin{split}
	\int_0^A \frac{w_1}{2}(u+Ru_R)\,dR+\varGamma_1 u(Z,A)&=\frac{w_1}{2} Ru|_{0}^A+\varGamma_1 u(Z,A)\\
	&=\Big(\frac{w_1}{2} A +\varGamma_1\Big) u(Z,A)=0,
	\end{split}
	\end{align}
where we have used the implicit equation \eqref{eq:implicit}. Then we proceed to find the function $v$ that minimizes $\mathcal{E}_2$. Upon integration by parts (see \ref{app:simp} for details), we can rewrite $\mathcal{E}_2$ in a compact form as
	\begin{align}\label{eq:E2d}
	\begin{split}
	\mathcal{E}_2=&\int_{-L}^L \Big[\int_0^A \Big(\frac{\zeta}{2} (v^{2}_R+R^2 a_\lambda^2 \lambda'(Z)^2)+\eta R a_\lambda\lambda'(Z)v_R\Big)R\,dR+\varGamma_2  \frac{A^3 a_\lambda^2\lambda'(Z)^2}{2\lambda}\Big]\,dZ\\
	& +\varGamma_2 A v(Z,A)|_{Z=-L}^{Z=L},
	\end{split}
	\end{align}
where the coefficients $\zeta$ and $\eta$ are given by
	\begin{align}\label{eq:bc}
	\zeta=\frac{\lambda w_2+a\varGamma_1/A }{\lambda^2-a^2},\quad \eta=\frac{a w_2+\lambda\varGamma_1/A}{\lambda^2-a^2}-\frac{\varGamma_{12}+\varGamma_{22}/a_\lambda}{A},
	\end{align}
in which $a_\lambda={da}/{d\lambda}$,  $\varGamma_{12}={\partial^2 \varGamma}/{\partial\lambda^\text{s}_1\partial\lambda^\text{s}_2}$, $\varGamma_{22}={\partial^2 \varGamma}/{{\partial(\lambda^{s}_2)}^2}$, and the derivatives are evaluated at $(a(Z),\lambda(Z))$.
By the completing square technique, we can rewrite the integrand in \eqref{eq:E2d} as
	\begin{align}
	\frac{\zeta}{2} (v^{2}_R+R^2 a_\lambda^2 \lambda'(Z)^2)+\eta R a_\lambda\lambda'(Z)v_R=\frac{\zeta}{2}\Big(v_R+\frac{\eta}{\zeta}  R a_\lambda\lambda'(Z)\Big)^2+\frac{\zeta^2-\eta^2}{2\zeta}R^2a_\lambda^2\lambda'(Z)^2.
	\end{align}
	The coefficient $\zeta$ of $v_R^2$ is positive, ensured by the convexity of the energy.  Hence, the energy functional \eqref{eq:EE} is minimized when
	\begin{align}\label{eq:vR}
	v_R=-\frac{\eta}{\zeta}a_\lambda R \lambda'(Z).
	\end{align}
	Integration of \eqref{eq:vR} subject to \eqref{eq:vcon} yields the optimal correction $v$ as
	\begin{align}\label{eq:vv}
	v(Z,R)=-\frac{\eta}{2\zeta}a_\lambda \lambda'(Z)\Big(R^2-\frac{1}{2}A^2\Big).
	\end{align}

\subsection{One-dimensional energy functional}
	
Upon substitution of \eqref{eq:vv} into \eqref{eq:EE}, after simplification, we obtain the final expression of the 1d energy functional
	\begin{align}\label{eq:1denergy}
	\mathcal{E}_\text{1d}[\lambda]=\int_{-L}^L \Big(\varPhi(\lambda)+\frac{1}{2}B(\lambda)\lambda'(Z)^2 \Big)\,dZ+C(\lambda)\lambda'(Z)|_{-L}^{L},
	\end{align}
	where $\varPhi(\lambda):=\varPhi(a(\lambda),\lambda)$, and the gradient moduli $B$ and $C$ are given by
	\begin{align}
	&B(\lambda)=A^4a_\lambda^2\Big(\frac{\zeta^2-\eta^2}{4\zeta}+ \frac{ \varGamma_2 }{\lambda A}\Big), \label{eq:BB}\\
	&C(\lambda)=-A^3 a_\lambda\varGamma_2\frac{\eta}{4 \zeta}.
	\end{align}
	It can be checked that our 1d model reproduces the 1d elasto-capillary model of \cite{LA2020a} by taking the surface energy function to be the form $\varGamma(\lambda_1^\text{s},\lambda_2^\text{s})=\gamma \lambda_1^\text{s} \lambda_2^\text{s}$, where $\gamma$ is the constant surface tension.

The equilibrium equation associated with the 1d energy functional can be derived via extremizing (\ref{eq:1denergy}) with respect to the function $\lambda(Z)$, yielding
	\begin{align}\label{eq:equi}
	&\varPhi'(\lambda)-\frac{1}{2}B'(\lambda)\lambda'(Z)^2-B(\lambda)\lambda''(Z)=0.
	\end{align}
	This equilibrium equation can be integrated once after multiplying the integration factor $\lambda'(Z)$,  resulting in the conservation law
	\begin{align}\label{eq:conservation}
	\varPhi(\lambda)-\frac{1}{2}B(\lambda)\lambda'(Z)^2=\text{constant},
	\end{align}
which is useful in finding solutions to the 1d model. 
	
We note that the energy functional \eqref{eq:1denergy} cannot be used directly to obtain the boundary conditions, as it is ill-posed due to the existence of the boundary term $C(\lambda)\lambda'(Z)|_{Z=-L}^{Z=L}$. This problem is rooted in the superficial term $\int_{-L}^L A\varGamma(\lambda_1^\text{s},\lambda_2^\text{s})\,dZ$  in the 3d variational problem \eqref{eq:E}. The variation of this term yields a singular surface stress concentrated at the outer edges of $Z=\pm L$ that cannot be balanced by the bulk stress. To cure the ill-posedness, we make use of the physical observation that the deformation away from the localized region is homogeneous. As a result, we may impose the essential boundary conditions $z(\pm L,A)=\pm l$ to supersede the natural boundary conditions, where $l$ is a constant that is tailored to satisfy the axial equilibrium of resultant forces.  This condition implies that $\lambda'(Z)=0$ at $Z=\pm L$ in view of \eqref{eq:asysol} and \eqref{eq:vv}, and the boundary term in $\mathcal{E}_\text{1d}$ vanishes as a result. With the boundary term neglected, the variation of \eqref{eq:1denergy} gives the correct boundary conditions for $\lambda(Z)$:
	\begin{align}\label{eq:LL}
	\lambda'(-L)=\lambda'(L)=0.
	\end{align}
	To the best of our knowledge, the above rigorous resolution of the ill-posedness of 1d model  has not been given in the literature.

\subsection{Solutions to the one-dimensional model}
	
	The last task is to find solutions to the 1d model.  Although in the previous derivation we have assumed that the cylinder is of finite length, localized instabilities like necking and bulging  can be analyzed more easily and very accurately by assuming that the cylinder is infinitely long \citep{WF2021}. This is because localized deformations decay exponentially towards the two ends, so that they are insensitive to the length of the cylinder.
	Thus we shall also assume that the cylinder is effectively infinitely long and replace the boundary condition \eqref{eq:LL} by the decaying boundary condition
	\begin{align}\label{eq:infty}
	\lim_{Z\rightarrow \pm\infty}{\lambda(Z)}=\lambda_\infty,
	\end{align}
	where $\lambda_\infty$ is a constant. A linear analysis shows that the solution to \eqref{eq:equi} satisfying \eqref{eq:infty} decays exponentially as $Z\to \infty$. It follows that  $\lim_{Z\rightarrow \pm\infty}{\lambda'(Z)}=0$. We further assume that the necking/bulging deformation is symmetric with respect to $Z=0$ such that
	\begin{equation}\label{eq:ic}
	\lambda(0)=\lambda_0,\quad \lambda'(0)=0.
	\end{equation}
	The initial value ``$\lambda_0$'' can be obtained by evaluating \eqref{eq:conservation} at $Z=0$ and $Z=\infty$, which yields
	\begin{align}\label{eq:iv}
	\varPhi(\lambda_0)=\varPhi(\lambda_\infty).
	\end{align}
	In particular,  a bifurcation occurs when \eqref{eq:iv} has a nontrivial solution  other than $\lambda_0=\lambda_\infty$. In view of the equilibrium equation $\varPhi'(\lambda_\infty)=0$, we see that the trivial solution $\lambda_0=\lambda_\infty$ is a double root of \eqref{eq:iv}. With the trivial solution factorized out,   we have from \eqref{eq:iv} that
	\begin{align}
	\lim_{\lambda_0\rightarrow\lambda_\infty}\frac{\varPhi(\lambda_0)-\varPhi(\lambda_\infty)}{(\lambda_0-\lambda_\infty)^2}=0.
	\end{align}
	By L'Hospital's rule, the bifurcation condition is given as
	\begin{align}
	\varPhi''(\lambda_{\text{cr}})=0,
	\end{align}
	where $\lambda_{\text{cr}}$ is the critical stretch at which a bifurcation arises from the homogeneous state. 
	
	Once $\lambda_0$ is determined,  the corresponding bifurcation solution can be obtained the solving the initial value problem
	\begin{align}\label{eq:1dmodel}
	&\varPhi'(\lambda)-\frac{1}{2}B'(\lambda)\lambda'(Z)^2-B(\lambda)\lambda''(Z)=0,\\
	&\lambda(0)=\lambda_0,\quad \lambda'(0)=0.
	\end{align}
	Such a second-order ODE system can be solved analytically using the method of separation of variables or numerically by the finite difference method. We refer to Refs. \cite{A2016} and \cite{yu2023} for details.

\section{Rayleigh-Ritz based finite-element scheme }\label{sec:finite-element method}

To validate the results of the 1d model, we further develop a FE numerical scheme to simulate 3d deformations of the cylinder due to the combined action of surface stress and axial force. The Rayleigh-Ritz method \citep{RR2014} is chosen to find the numerical solutions of the 3d nonlinear model. It is a direct method for minimizing a given functional, yielding a solution to the variational problem without  having to solve the associated Euler-Lagrange equation. The simplicity and flexibility of the Rayleigh-Ritz method allow us to trace the bifurcation starting from an approximate 1d solution instead of the undeformed state. Because of this, no  imperfections are needed  in our numerical simulations and the computation results are  accurate even in the vicinity of the bifurcation point, as shown in loading curves (Figs. \ref{fig:solution12a}, \ref{fig:solution22} and \ref{fig:solution42}) in the next section. This is in contrast with the numerical simulations using FE software  \citep{TC2015}, which requires an initial imperfection to trigger the instability and yields results having incorrect behavior near the bifurcation threshold; see the discussion in \cite{LA2018}.

To implement the numerical scheme, we start from the 3d  potential energy functional \eqref{eq:E}. As pointed out in the previous section, the variational problem associated with \eqref{eq:E} is ill-posed due to the presence of the superficial term $\int_{-L}^{L} A\varGamma\,dZ$. This ill-posedness renders it challenging to impose the boundary conditions at the cylinder, especially in force-controlled problems. Here, we propose a new approach to resolve this issue.
	
To keeps the numerical scheme as general as possible, we consider  the loading scenario that the axial force $N$ is prescribed. Other loading scenarios can be addressed by tailoring this general case. Based on the argument in the previous section, we  cure the ill-posedness by imposing the essential boundary conditions $z(\pm L,A)=\pm l$, where  $l$ is an adjustable constant, calibrated to meet the axial equilibrium of resultants. This essential boundary conditions can be incorporated into the energy functional \eqref{eq:E}  by the Lagrange multiplier method.  When the bulk material is incompressible, a penalty term  $-p(J-1)$ should also be added to the energy functional, where $p$ is the Lagrange multiplier enforcing the incompressibility constraint and $J=\text{det}(\bm{F})$. By doing so, we are led to consider minimizing the following energy functional
	\begin{align}\label{eq:E3}
	\tilde{\mathcal{E}}[z,r,p,q]=\int_{0}^{L}\Big[\int_0^A \Big(W-p(J-1) \Big)R\,dR+A\varGamma \Big]\,dZ-\int_0^A q\Big(z(L,R)-l\Big)\,dR,
	\end{align}
	where the Lagrange multiplier $q$ represents the force per unit reference cross-section area at $Z=L$ times  $R$ and satisfies the axial resultant equilibrium
	\begin{align}\label{eq:resultant}
	2\pi  \int_0^A q \,dR=N\pi A^2.
	\end{align}
Note that we have assumed that the 3d  deformation is symmetric with respect to $Z=0$.

A four-node bilinear quadrilateral axisymmetric element (Q4) is employed in the numerical calculations. We partition the domain $[0,L]\times [0,A]$ into $n\times m$ equal squares. On each of the squares, the position functions  $z(Z,R)$, $a(Z,R)=r(Z,R)/R$ and Lagrange multiplier $p(Z,R)$ are approximated by bilinear interpolation in terms of their nodal values, and linear interpolation is used for the Lagrange multiplier $q(R)$. Here, introducing the function $a$ is to remove the singularity of the transversal stretch $r/R$ at $R=0$. For example, on a typical square $[ih,(i+1)h]\times [jh,(j+1)h]$ where $h=L/n$ is the mesh size, the functions $z(Z,R)$ and $q(R)$ are approximated by
	\begin{align}
	&z(Z,R)=z_{i,j}N_1(\xi,\theta)+z_{i+1,j}N_2(\xi,\theta)+z_{i+1,j+1}N_3(\xi,\theta)+z_{i,j+1}N_4(\xi,\theta),\\
	&q(R)=q_j+\frac{q_{j+1}-q_j}{h}(R-jh),
	\end{align}
	where $\xi={(Z-ih)}/{h}$ and $\theta={(R-jh)}/{h}$ are the local coordinates, $z_{i,j}$ and $q_j$ denote respectively the numerical approximation of $z(ih,jh)$ and $q(jh)$, and $N_k(\xi,\theta)$, $1\leq k\leq 4$ are the shape functions defined by
	\begin{equation}
	\begin{aligned}
	&N_1(\xi,\theta) = (\xi-1)(\theta-1), \quad N_2(\xi,\theta) = \xi(1-\theta), \\
	&N_3(\xi,\theta) = \xi\theta, \hspace{6em} N_4(\xi,\theta)= (1-\xi)\theta.
	\end{aligned}
	\end{equation}

Substituting these approximations into \eqref{eq:E3} and calculating the integration on each square using a four-point Gaussian quadrature rule, we obtain a discrete energy of the form
	\begin{align}\label{eq:E4}
	\tilde{\mathcal{E}}[z,r,p,q]=\tilde{\mathcal{E}}(z_{i,j},a_{i,j},p_{i,j},q_{j}),\quad 0\leq i\leq n,\ 0\leq j\leq m.
	\end{align}
	From the stationary condition of \eqref{eq:E4} as well as \eqref{eq:resultant} and elimination of $q_j$, we get $3(n+1)(m+1)+1$ equations for the corresponding unknowns $z_{i,j}$, $a_{i,j}$, $p_{i,j}$, $0\leq i\leq n,\ 0\leq j\leq m$ and $l$,
	\begin{align}\label{eq:ae}
	\begin{split}
	&z_{0,j}=0,\quad \frac{\partial\tilde{\mathcal{E}}}{\partial z_{i,j}}=0,\ z_{n,j}=l,\quad 1\leq i\leq n-1,\ 0\leq j\leq m,\\
	&\frac{\partial\tilde{\mathcal{E}}}{\partial a_{i,j}}=0,\quad \frac{\partial\tilde{\mathcal{E}}}{\partial p_{i,j}}=0,\quad 0\leq i\leq n,\ 0\leq j\leq m,\\
	&\sum_{j=0}^m \frac{\partial\tilde{\mathcal{E}}}{\partial z_{n,j}}-\frac{1}{2}NA^2=0.
	\end{split}
	\end{align}
Note that the symmetric boundary condition $z_{0,j}=0$ is imposed to avoid the singularity of the tangent stiffness matrix of the system. In the original formulation \eqref{eq:E}, the end force is imposed in a pointwise manner ${\partial \tilde{\mathcal{E}}}/{\partial z_{n,j}}=N R_j$ with $R_j=jh$, $j=0,1,\dots,n$, which thus cannot support singular stress distribution. Our new formulation, in particular $\eqref{eq:ae}_3$, relaxes the end condition to a resultant condition, which results in the correct stress distribution at the end and remedies the convergence of the numerical solutions, as shown in Figs. \ref{fig:solution11b} and \ref{fig:solution22b}.  A detailed comparison of the results obtained from these two formulations is provided in the supplementary material.

The system of algebraic equations \eqref{eq:ae} can be solved using the Newton-Raphson method, if a suitable initial guess is given.  The solution of the 1d model offers a good candidate for the initial guess. Once a solution in the post-bifurcation regime is found,  the solution to the entire post-bifurcation regime can be traced throughout by iteratively using the solution at the previous step as the initial guess for the current step.  We have implemented the above numerical scheme in the symbolic computation software {\it Mathematica}. The large system of algebraic equations is defined in Mathematica with the built-in command \texttt{Table} and is solved using the built-in solver \texttt{FindRoot} which checks numerical stability automatically. To ensure the convergence of the FE calculations, we start from a coarse mesh and gradually refine the mesh. The relative error, calculated as the relative difference between the solution from a finer mesh and the current solution,  is monitored during the process. The solution is considered to be convergent if the relative error reaches less than $10^{-3}$. The simulation parameters and calculation time for each figures are summarized in Table 1 of the supplementary material. A single numerical solution can typically be obtained in a few seconds on a personal computer, which shows the considerable efficiency of the FE scheme.

\section{Results and discussion}\label{sec:analysis}
	
This section presents the efficiency of 1d model via comparisons with the 3d FE simulations and to explore how the surface elasticity parameters affect the localized necking/bulging in the cylinder based on the theoretical results. 
	
For numerical illustration, we consider that the bulk is made of the incompressible neo-Hookean material \cite{K2012}, with the strain energy function
	\begin{align}\label{eq:W}
	W=\frac{\mu}{2}(I_1-3),
	\end{align}
where $\mu$ is the shear modulus and $I_1=\lambda_1^2+\lambda_2^2+\lambda_3^2$ is the first principal invariant. The surface energy function is assumed to take the form \citep{BD2023}
	\begin{align}\label{eq:Gamma}
	\varGamma=\gamma J_\text{s}+\frac{\mu_\text{s}}{2}(I_1^\text{s}-2-2\ln J_\text{s})+\frac{\mu_\text{s}\nu_\text{s}}{1-\nu_\text{s}}\Big(\frac{1}{2}(J_\text{s}^2-1)-\ln J_\text{s}\Big),
	\end{align}
where $\gamma$ is the (residual) surface tension, $\mu_\text{s}$ is the shear modulus of the surface,  $\nu_\text{s}$ is the surface Possion's ratio, $I_1^\text{s}=({\lambda_1^\text{s}})^2+({\lambda_2^\text{s}})^2$ and $J_\text{s}=\lambda_1^\text{s}\lambda_2^\text{s}$. Note that in surface elasticity, since the second invariant $I_2^\text{s}$  is fundamentally linked to $J_\text{s}$  by $I_2^\text{s}=\det{(\bm{F}_\text{s}^T\bm{F}_\text{s})}=J_\text{s}^2$ \cite{Hui2020},  the dependence of the surface energy on  $J_\text{s}$ inherently incorporates $I_2^\text{s}$. With the above bulk and surface material models, the expression of the 1d model \eqref{eq:1dmodel} can be specified (given in \ref{app:neo-Hookean}). Note that the 1d model is applicable when the bulk material is compressible, and the current incompressible case is obtained as the limiting case of the compressible one (e.g, letting the bulk Poisson's ratio  go to $1/2$). 
	
We remark that the neo-Hookean model is selected for the bulk material because it provides a simplest framework to disentangle surface effects from complex bulk nonlinearities. The bulk nonlinear effects may also modulate the elasto-capillary interactions. For instance, the effect of the second invariant $I_2$ can make the necking/bulging more evident but does not influence the qualitative trends of the bifurcation (see Subsection S1.3 of the supplementary material). The viscoelastic effects can significantly alter the dynamics of finite-wavelength instability (e.g., reducing growth rates and shifting critical wavenumbers) while preserving the static stability boundary \cite{B2018,T2021}.  Material anisotropy could introduce directional dependence in the instability, potentially altering the symmetry of the resulting patterns. Additionally, inertial effects participate in energy conversion, influence dynamic evolution, and together with surface stress and elastic forces determine deformation rate, mode, final morphology \cite{P2021,TB2025}. Our bulk and surface models intentionally assume isotropy and hyperelasticity, neglecting viscoelastic, anisotropic, and inertial effects. These extensions warrant future study. Nevertheless, despite these simplifications, the model provides a foundational framework for probing strain-dependent surface effects. 
	
For convenience, we use $\mu$ and $A$ as the  stiffness and characteristic length of the system and introduce the following dimensionless quantities
	\begin{align}
	\bar{N}=\frac{N}{\mu},\quad \bar{\gamma}=\frac{\gamma}{\mu A},\quad \bar{\mu}_s=\frac{\mu_s}{\mu A}.
	\end{align}
	The quantities $\bar{\gamma}$ and $\bar{\mu}_\text{s}$ are referred to as the {\it elasto-capillary number} and {\it surface-elasticity number}, respectively \citep{BD2023}. In the sequel, either the dimensionless axial force $\bar{N}$ or the elasto-capillary number $\bar{\gamma}$ is treated as bifurcation parameters, while the surface-elasticity number $\bar{\mu}_\text{s}$ and the surface Poisson ratio $\nu_\text{s}$ are set as influential parameters with fixed values. Three typical loading scenarios \citep{Fu2021,DJ2022,Emery2023} are considered: (i) fixing $\bar{\gamma}$ and varying $\bar{N}$; (ii) fixing $\bar{N}$ and varying $\bar{\gamma}$; (iii) fixing the two ends (i.e. fixing the average stretch) and varying  $\bar{\gamma}$.

\subsection{Varying axial force $\bar{N}$ at fixed elasto-capillary number $\bar{\gamma}$}
	
For the scenario of fixed surface parameters, when the axial force $\bar{N}$ increases  from zero to a critical value $\bar{N}_{\text{cr}}$, necking will initiate (assuming  it occurs at $Z=0$). Simultaneously, the axial force begins to drop from $\bar{N}_{\text{cr}}$. The necking grows continuously until the axial force drops to the Maxwell stress $\bar{N}_\text{M}$. At this stage, the necking amplitude reaches its maximum. The necking begins to propagate steadily to the two ends of the cylinder and the stress stays at the Maxwell level. The effects of surface elasticity number $\bar{\mu}_\text{s}$, surface Poisson's ratio $\nu_\text{s}$ on this necking behaviour are analyzed in the sequel. We remark that the effect of elasto-capillary number $\bar{\gamma}$ is not our main concern currently, so we present the results for $\bar{\gamma}$ in the supplementary material.
	
\subsubsection{Effect of surface elasticity}
	\begin{figure}[ht!]
		\centering
		\includegraphics[width=\linewidth]{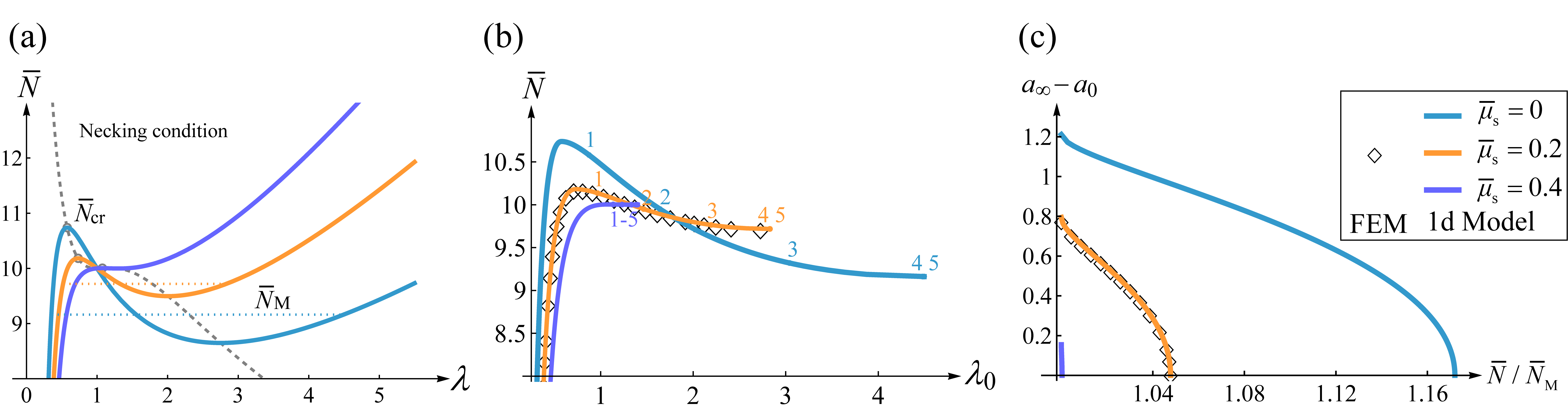}
		\caption{Homogeneous and inhomogeneous responses for the cases of fixed  $\bar{\mu}_\text{s}=0$, $0.2$, $0.4$ respectively, $\bar{\nu}_\text{s}=0.2$ and $\bar{\gamma}=10$. (a) Homogeneous responses of the axial stress $\bar{N}$ varying with the stretch $\lambda$. (b) Dependence of the axial stress $\bar{N}$ on the stretch $\lambda_0=\lambda(0)$ at the necking starting point. (c) Necking amplitudes $a_\infty-a_0$ varying with the ratio of the axial stress to the Maxwell stress $\bar{N}/\bar{N}_\text{M}$. The gray dashed line in (a) represents the necking condition. The dotted parallel lines in (a) represent the corresponding Maxwell plateaus $\bar{N}_\text{M}$. The scattered hollow dots in (b) and (c) are the 3d FE results  for $\bar{\mu}_\text{s}=0.2$. The solid points marked $1$-$5$ in (b) denote the different post-bifurcation stages. The deformations of the   cylinder at these stages are presented in Fig. \ref{fig:solution11b}. }\label{fig:solution12a}
	\end{figure}
	
	\begin{figure}[ht!]
		\centering
		\includegraphics[width=\linewidth]{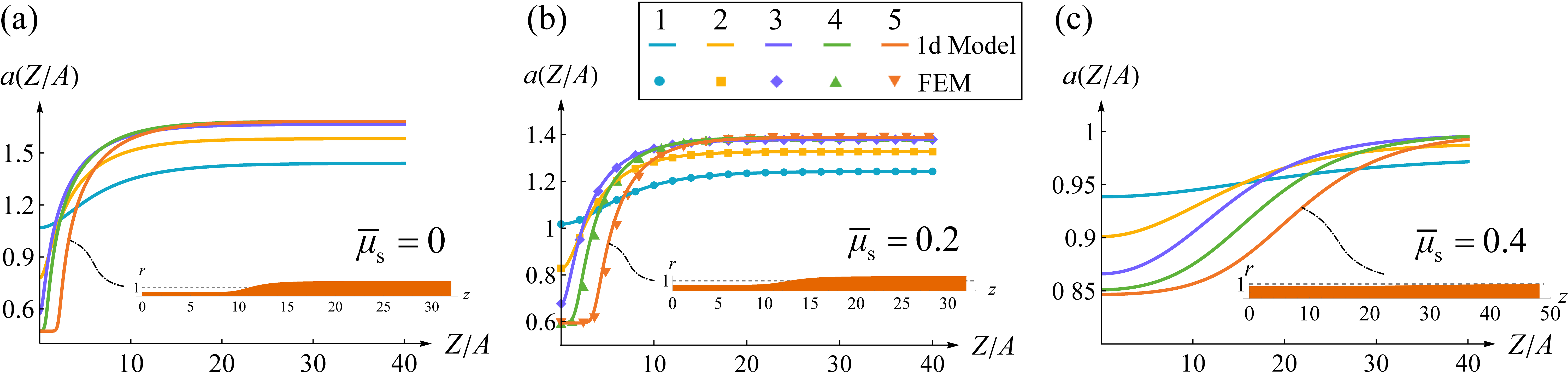}
		\caption{Necking solutions for the radial stretch $a(L/A)$ at the five post-bifurcation stages marked in Fig. \ref{fig:solution12a}(b) for the three different surface-elasticity numbers: (a) $\bar{\mu}_\text{s}=0$, (b) $\bar{\mu}_\text{s}=0.2$ and (c) $\bar{\mu}_\text{s}=0.4$. The orange-colored  section displays the morphology of the deformed cylinder  at the final stage of bifurcation, which records the maximum necking amplitude.}\label{fig:solution11b}
	\end{figure}
	
We first explore the impact of the elasto-capillary number $\bar{\mu}_\text{s}$ by studying the case of $\bar{\mu}_\text{s}=0$, $0.2$, $0.4$ respectively, $\bar{\nu}_\text{s}=0.2$ and $\bar{\gamma}=10$. Fig. \ref{fig:solution12a}(a) presents the homogeneous $\bar{N}$-$\lambda$ responses, which is determined by solving $\varPhi'(\lambda)=0$ for $\bar{N}$. Fig. \ref{fig:solution12a}(b) shows the asymptotic solutions of the 1d model for $\bar{N}$ varying with the stretch at $Z=0$ in the entire post-bifurcation regime. The necking amplitudes varying with ratio $\bar{N}/\bar{N}_M$ are shown in Fig. \ref{fig:solution12a}(c). Moreover, we take five points on each curve in Fig. \ref{fig:solution12a}(b) to show the necking evolutions; see Fig. \ref{fig:solution11b}. Note that the fourth and fifth points almost overlap each other.
In Fig. \ref{fig:solution12a}(b) and (c), we also present the numerical results of the FE simulations for the case of $\bar{\mu}_\text{s}=0.2$ (see the scattered hollow points). The FE solutions of $a(Z/A)$ corresponding to the five marked points $1$-$5$ are shown in  Fig. \ref{fig:solution11b}(b), for which the corresponding necking morphologies are shown in Fig. S1 of the supplementary material. It is seen that the results of the 1d model agree very well with the results of FE simulations even in the finial stage of necking growth, which demonstrates the high efficiency of the derived 1d model.

We see that the  curves of the homogeneous responses intersect at $\lambda=1$. It implies that when the cylinder is not stretched, the axial force is independent of surface elasticity (as well as surface compressibility). Another feature is that for larger $\bar{\mu}_\text{s}$, the necking can be triggered at a lower $\bar{N}_\text{I}$, but propagates at a higher Maxwell plateau $\bar{N}_\text{M}$. Consequently, the growth of the necking is more rapid and less apparent for larger $\bar{\mu}_\text{s}$. This can be further observed from the necking morphologies at the final stage of necking growth (corresponding to stage 5) for each case of $\bar{\mu}_\text{s}$ as shown in Fig. \ref{fig:solution11b}. In particular, when $\bar{\mu}_\text{s}$ only reaches $0.4$, the necking is nearly imperceptible compared with the case of zero surface elasticity. The necking instability in the cylinder structure under the present loading scenario (with $\bar{\gamma}=10$, $\nu_\text{s}=0.2$) can be effectively mitigated by adjusting the surface-elasticity number within the range of $0$ to $0.4$.
	
\subsubsection{Effect of surface compressibility}

\begin{figure}[ht!]
		\centering
		\includegraphics[width=\linewidth]{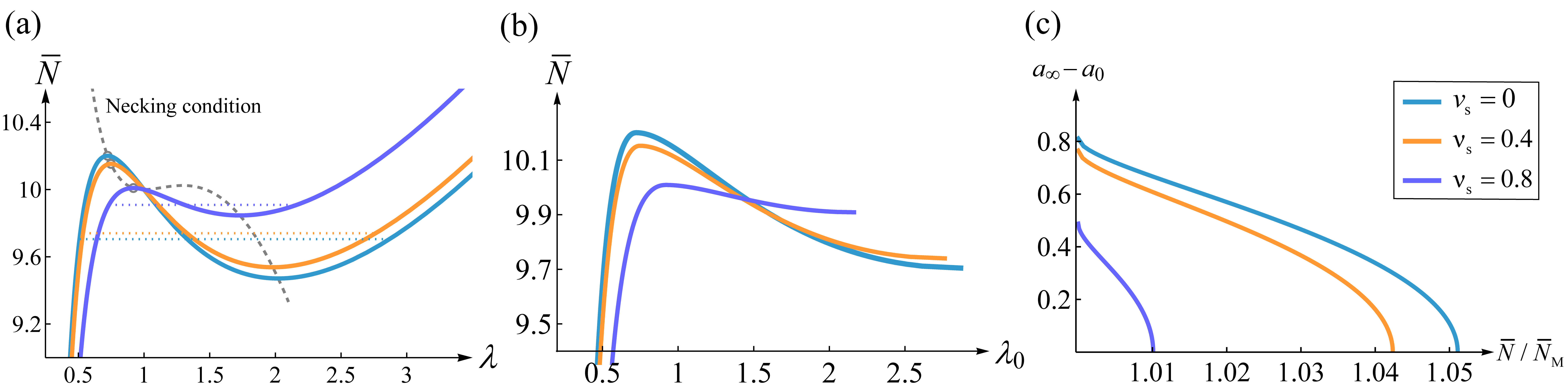}
		\caption{Homogeneous and inhomogeneous responses for the cases of fixed $\bar{\gamma}=10$, $\bar{\mu}_\text{s}=0.2$ and $\nu_\text{s}=0$, $0.4$, $0.8$ respectively. (a) Homogeneous responses for the axial stress $\bar{N}$ varying with the stretch $\lambda$. (b) Dependence of the axial stress $\bar{N}$ on the stretch $\lambda_0=\lambda(0)$ at the necking starting point. (c) Necking amplitudes $a_\infty-a_0$ varying with the ratio of the axial stress and the Maxwell stress $\bar{N}/\bar{N}_\text{M}$. The gray dashed line in (a) represents the necking condition. The dotted parallel lines in (a) represent the corresponding Maxwell plateaus $\bar{N}_\text{M}$. }\label{fig:solution13a}
	\end{figure}
	
	The effect of the surface compressibility depicted by the surface Poisson's ratio $\nu_\text{s}$ is also investigated (see Fig. \ref{fig:solution13a}). Note that a surface with higher $\nu_\text{s}$ represents higher incompressibility. Akin to  $\bar{\mu}_\text{s}$, increasing $\nu_\text{s}$ can advance the necking initiation but reduce the amplitude. It is observed that the necking behaviors for $\nu_\text{s}=0$ and $0.4$ do not differ much in contrast to the difference for $\nu_\text{s}=0.4$ and $0.8$. This suggests that the necking instability can be more effectively controlled by adjusting the surface Poisson's ratio within the range of high values.

\subsubsection{Sensitivity analysis of surface parameters}

To  further explore how uncertainties in the surface parameters impact the PR instability, we conduct a sensitivity analysis of the surface parameters by using the Morris screening method  \cite{M1991,C2007}. The Morris method is a global sensitivity analysis technique that evaluates the influence of input parameters on model outputs through elementary effects. It is particularly efficient for screening important parameters in complex models. 
		
The procedure of the sensitivity analysis is as follows. We restrict the three surface parameters to the intervals: $\bar{\gamma}\in[6,20]$, $\bar{\mu}_\text{s}\in[0,0.5]$ and $\nu_\text{s}\in[0,0.9]$ and follow the algorithm of the Morris method to obtain a discrete parameter set. Each input parameter $x_i$ ($x_1=\bar{\gamma}$, $x_2=\bar{\mu}_\text{s}$ or $x_3=\nu_\text{s}$) is divided into $p=4$ levels. $R=10$ random trajectories are generated in parameter space. Then, we compute the elementary effect $EE_i^{(j)}$ by evaluating the maximum amplitude (the amplitude in the final bifurcation stage) $\Lambda_\text{max}=\max\{|a_\infty-a_0|\}$ at each trajectory point pair 
		\begin{equation}\begin{aligned}
		&      EE_1^{(j)} = \frac{\Lambda_\text{max}|_{(\bar{\gamma},\bar{\mu}_\text{s},\nu_\text{s})=(x_1^{(j)},x_2,x_3)} - \Lambda_\text{max}|_{(\bar{\gamma},\bar{\mu}_\text{s},\nu_\text{s})=(x_1,x_2,x_3)}}{\Delta},
		\\&EE_2^{(j)} = \frac{\Lambda_\text{max}|_{(\bar{\gamma},\bar{\mu}_\text{s},\nu_\text{s})=(x_1,x_2^{(j)},x_3)} - \Lambda_\text{max}|_{(\bar{\gamma},\bar{\mu}_\text{s},\nu_\text{s})=(x_1,x_2,x_3)}}{\Delta},
		\\&EE_3^{(j)} = \frac{\Lambda_\text{max}|_{(\bar{\gamma},\bar{\mu}_\text{s},\nu_\text{s})=(x_1,x_2,x_3^{(j)})} - \Lambda_\text{max}|_{(\bar{\gamma},\bar{\mu}_\text{s},\nu_\text{s})=(x_1,x_2,x_3)}}{\Delta}.
		\end{aligned}
		\end{equation}
	
The trajectory point pairs are generated randomly. Note that the amplitude is set to $0$ if at the parameter set, PR instability does not happen in the system. 
		
Two statistics are calculated for each parameter: the mean absolute elementary effect
		\begin{equation}
		E_i = \frac{1}{R}\sum_{j=1}^R |EE_i^{(j)}|
		\end{equation}
		and    
		the standard deviation of elementary effects
		\begin{equation}
		\sigma_i = \sqrt{\frac{1}{R-1}\sum_{j=1}^R (EE_i^{(j)} - E_i)^2}.
		\end{equation}
		The mean absolute elementary effect $E_i$ measures the overall influence of the parameter $x_i$, and the standard deviation $\sigma_i$ indicates the nonlinearity and interaction effects. The results are presented in Table \ref{parameter}.
		
		\begin{table}[h]
			\centering
			\caption{Sensitivity metrics for surface parameters}\label{parameter}
			\begin{tabular}{cccc}
				\hline
				Parameter & $E_i$ & $\sigma_i$ & Influence characteristics \\
				\hline
				$\bar{\gamma}$ & 0.12 & 0.04 & Linear parameter (weak) \\
				$\bar{\mu}_\text{s}$ & 2.31 & 1.33& Nonlinear/Interactive parameter (strong) \\
				$\nu_\text{s}$ & 0.03 & 0.07 & Negligible parameter  \\
				\hline
			\end{tabular}
		\end{table}
		
The results reveal distinct influence patterns among the three surface parameters. The surface-elasticity number $\bar{\mu}_\text{s}$ exhibits the strongest influence on system behavior, as evidenced by high mean absolute effect ($E_2 = 2.31$), indicating substantial overall impact; large standard deviation ($\sigma_2 = 1.33$), suggesting either strong nonlinear effects in the parameter's influence or significant interactions with other parameters; elementary effects ranging from $-3.88$ to $0$, showing $\bar{\mu}_\text{s}$ can both stabilize and destabilize the system depending on its value. The elasto-capillary number $\bar{\gamma}$ demonstrates relatively weak yet consistent influence: a small positive mean effect  ($E_1 = 0.12$) across all elementary effects; low variability ($\sigma_1 = 0.04$), suggesting nearly linear influence on system response and minimal interaction with other parameters;  all elementary effects fall within $[0.08, 0.17]$, indicating reliable proportional impact. The surface Poisson's ratio $\nu_\text{s}$ shows minimal influence: very small mean effect ($E_3 = 0.03$) and sparse nonzero elementary effects ($-0.20$ to $0$);
		
The above analysis shows that uncertainties in surface parameters impact the model's predictions asymmetrically: the surface-elasticity number $\bar{\mu}_\text{s}$ is the primary driver of variability and nonlinearity, the elasto-capillary number $\bar{\gamma}$ introduces minor linear shifts, and the surface Poisson's ratio $\nu_\text{s}$ is relatively irrelevant. Thus, efforts to quantify/control uncertainties should focus on the surface elasticity for robust predictions.

\subsection{Varying elasto-capillary number $\bar{\gamma}$ at fixed axial force $\bar{N}$}
	
With the fixed axial force $\bar{N}$ applied, the cylinder is first stretched to some value. Then the elasto-capillary number $\bar{\gamma}$ increases from zero. The shrinkage effect of $\bar{\gamma}$ leads to a decrease of the stretch. For $\bar{N}$  large enough, bulging can nucleate (at the center of the cylinder) once $\bar{\gamma}$ reaches a critical value $\bar{\gamma}_{\text{cr}}$. Subsequently, $\bar{\gamma}$ will drop from the initiation value $\bar{\gamma}_{\text{cr}}$ to the Maxwell value $\bar{\gamma}_\text{M}$ and the bulging grows simultaneously to reach its maximum amplitude. After this, the bulging region extends to the two ends of the cylinder with $\bar{\gamma}$ kept at the Maxwell level. In the following, we analyze the effects of surface-elasticity number $\bar{\mu}_\text{s}$ and surface compressibility parameter $\nu_\text{s}$ on the bulging response respectively.  Note that the effect of axial force $\bar{N}$ is analyzed  in the supplementary material.
	\begin{figure}[h]
		\centering
		\includegraphics[width=\linewidth]{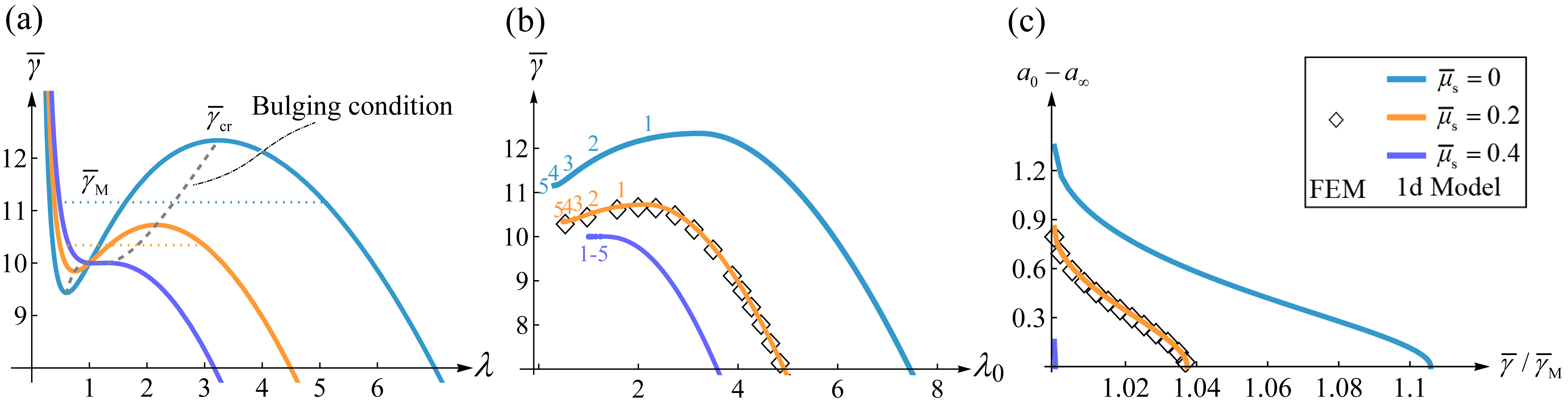}
		\caption{Homogeneous and inhomogeneous responses for the cases of fixed  $\bar{N}=10$, $\bar{\mu}_\text{s}=0$, $0.2$, $0.4$ respectively, and $\nu_\text{s}=0.2$. (a) Homogeneous responses of the elaso-capillary number $\bar{\gamma}$ varying with the stretch $\lambda$. (b) Dependence of the $\bar{\gamma}$ on the stretch $\lambda_0=\lambda(0)$ at the bulging starting point. (c) Bulging amplitudes $a_0-a_\infty$ varying with ratio of the surface tension and its Maxwell level $\bar{\gamma}/\bar{\gamma}_\text{M}$. The gray dashed line in (a) represents the bulging condition. The dotted parallel lines in (a) represent the corresponding Maxwell plateaus $\bar{\gamma}_\text{M}$. The scattered hollow dots in (b) and (c) are the 3d FE results  for $\bar{\mu}_\text{s}=0.2$. The solid points marked $1$-$5$ in (b) denote the different post-bifurcation stages. The deformations of the cylinder at these stages are presented in Fig. \ref{fig:solution22b}. }\label{fig:solution22}
	\end{figure}
	
\begin{figure}[ht!]
		\centering
		\includegraphics[width=\linewidth]{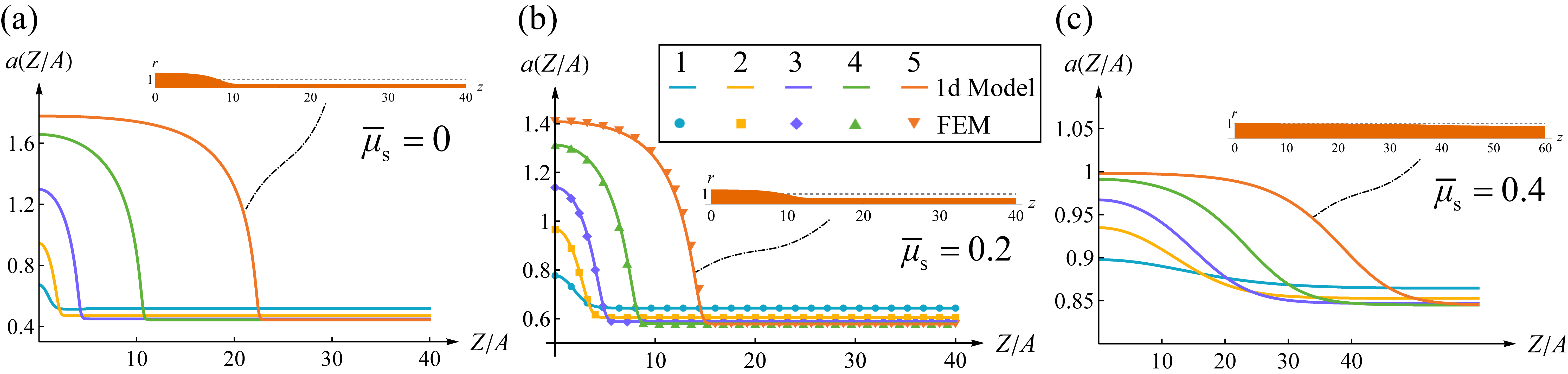}
		\caption{Bulging solutions for the radial stretch $a(L/A)$ at the five post-bifurcation stages marked in Fig. \ref{fig:solution22}(b) for the three different surface-elasticity numbers: (a) $\bar{\mu}_\text{s}=0$, (b) $\bar{\mu}_\text{s}=0.2$ and (c) $\bar{\mu}_\text{s}=0.4$. The orange-colored  section displays the morphology of the deformed cylinder  at the final stage of bifurcation, which records the maximum bulging amplitude. }\label{fig:solution22b}
	\end{figure}

The cases with fixed $\bar{N}=10$, $\bar{\mu}_\text{s}=0$, $0.2$, $0.4$ respectively, and $\nu_\text{s}=0.2$ is first inspected to explore the effect of surface elasticity. Fig. \ref{fig:solution22}(a) shows the homogeneous $\bar{\gamma}$-$\lambda$ responses. The $\bar{\gamma}$-$\lambda_0$ responses and the bulging amplitude varying with $\bar{\gamma}$ are presented in Fig. \ref{fig:solution22}(b) and (c) respectively. The 1d model solutions for $a(Z/A)$ at the five stages marked in Fig. \ref{fig:solution22}(b) are exhibited in Fig. \ref{fig:solution22b}. The corresponding 3d FE results are also presented in these figures for comparison. It is seen that for this loading scenario, the 1d model also shows excellent consistency with the 3d FE simulations. The necking evolution patterns corresponding to the five deformation stages are shown in the supplementary material.

Different from the previous scenario, increasing $\bar{\mu}_\text{s}$ can prevent both the occurrence and growth of bulging instability. In particular, with fixed $\bar{N}=10$ and $\nu_\text{s}=0.2$, the amplitude and the propagation of the bulging for a surface-stiffening effect quantized by $\bar{\mu}_\text{s}=0.4$ is almost undetectable (see the bulging morphology shown in Fig. \ref{fig:solution22b}(c)). This implies that the bulging instability can also be avoided by raising the surface elasticity.
	
\begin{figure}
		\centering
		\includegraphics[width=\linewidth]{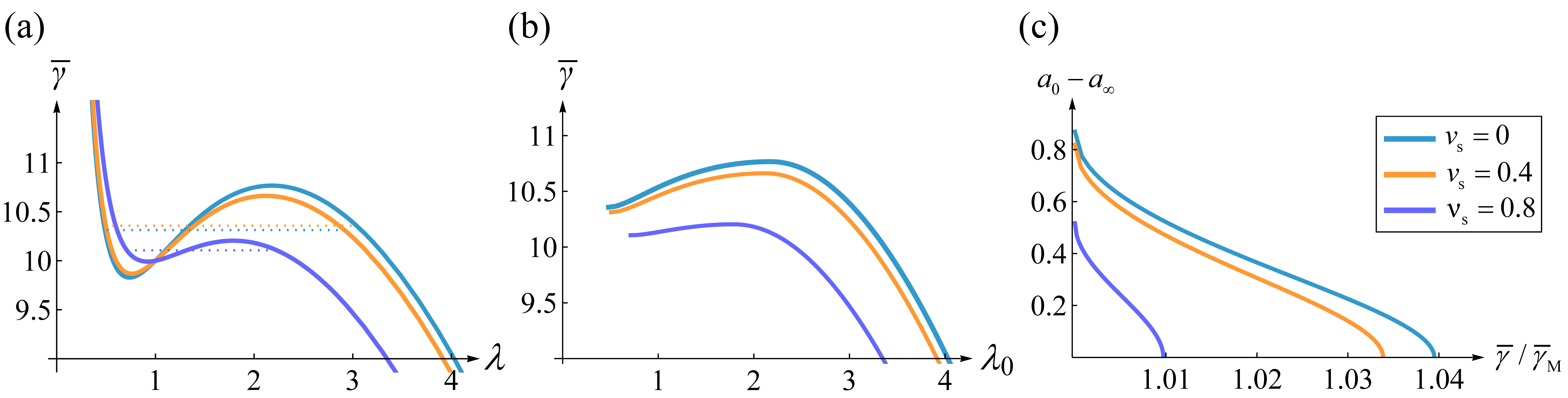}
		\caption{Homogeneous and inhomogeneous responses for the cases of fixed  $\bar{N}=10$, $\bar{\mu}_\text{s}=0.2$, and $\nu_\text{s}=0,~0.4,~0.8$ respectively. (a) Homogeneous responses for the elasto-capillary number $\bar{\gamma}$ varying with the stretch $\lambda$. (b) Dependence of $\bar{\gamma}$ on the stretch $\lambda_0=\lambda(0)$ at bulging starting point. (c) Bulging amplitudes $a_0-a_\infty$ varying with the ratio of the surface tension to its Maxwell level $\bar{\gamma}/\bar{\gamma}_\text{M}$. The dotted parallel lines in (a) represent the corresponding Maxwell plateaus $\bar{\gamma}_\text{M}$. }\label{fig:solution23}
\end{figure}
The impact of the surface Poisson's ratio in the second loading scenario is shown by Fig. \ref{fig:solution23}.  The effect of $\nu_\text{s}$ is akin to that of $\bar{\mu}_\text{s}$. Both the bifurcation condition and the amplitude can be reduced by increasing the surface incompressibility.
	
\subsection{Varying elasto-capillary number $\bar{\gamma}$ at fixed ends}

In this scenario, the cylinder is first stretched to a given length $2l$, corresponding to an applied stretch $\hat{\lambda}=l/L$, and then its two ends are fixed to prevent any further displacements. By analyzing the bifurcation condition $\varPhi''(\lambda)=0$ we find that $\bar{\gamma}$ has a minimum at a specific stretch $\lambda_{\text{min}}$, such that if the applied stretch $\hat{\lambda}<\lambda_{\text{min}}$,  necking will occur as $\bar{\gamma}$ increases to the critical value $\bar{\gamma}_\text{cr}$, whilst bulging will initiate instead if $\hat{\lambda}>\lambda_{\text{min}}$. This situation is similar to that for the elasto-capillary problem with constant surface tension, which was discovered by Fu et al. \cite{Fu2021}. With $\bar{\gamma}$ increasing from $\bar{\gamma}_\text{cr}$, the necking/bulging amplitude increases continuously and the phase interface propagates from the center to the two ends of the cylinder. For the sake of conciseness, in the following we discuss the results for a $\hat{\lambda}$ such that bulging can initiate. The necking situations are given in the supplementary material.

\subsubsection{Effect of surface-elasticity number}

The influence of surface-elasticity number in this loading scenario is investigated by considering $\bar{\mu}_\text{s}=0$, $0.2$, $0.4$ respectively, $\nu_\text{s}=0.2$ and $L/A=40$. For these three parameter sets,  the critical elasto-capillary number takes its minimum at the stretch $\lambda_\text{min}=1.2599$, $1.2134$, $1.1853$, respectively. Thus, we take $\hat{\lambda}=1.5$ to study the bulging instability. 
	
	\begin{figure}[ht!]
		\centering
		\includegraphics[width=1\linewidth]{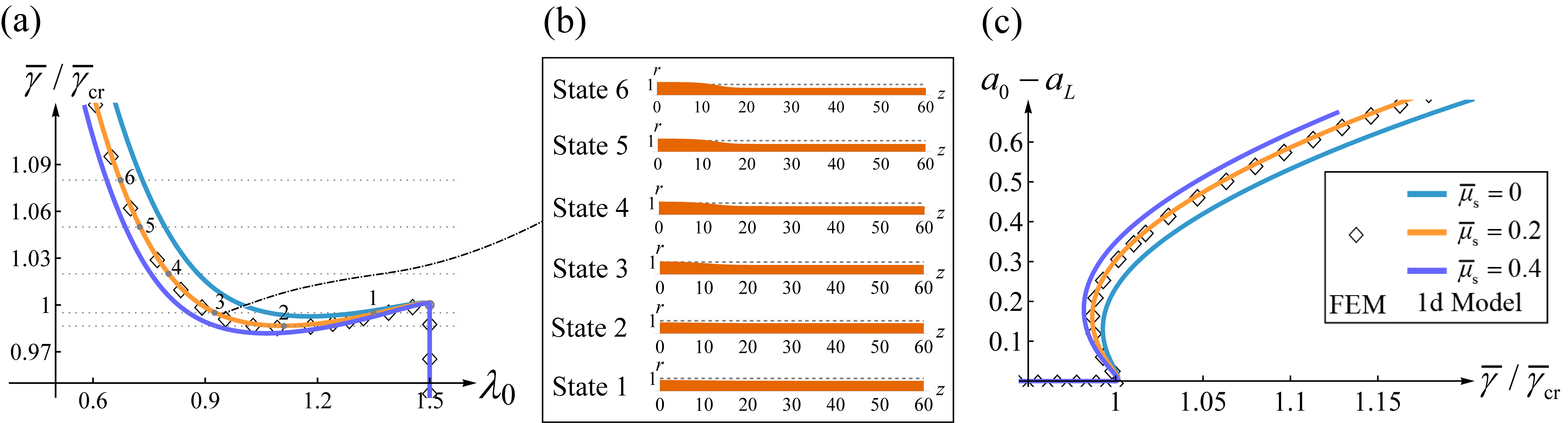}
		\caption{Bulging responses for the cases of fixed $\hat{\lambda}=1.5$ and $\bar{\mu}_\text{s}=0$, $0.2$, $0.4$ respectively, $\nu_\text{s}=0.2$ and $L/A=40$. (a) Dependence of the elaso-capillary number $\bar{\gamma}$ on the stretch $\lambda_0=\lambda(0)$ at the bulging starting point. The solid points marked $1$-$5$ in the curve for $\bar{\mu}_\text{s}=0.2$ denote different bulging stages, at which the cylinder morphologies are displayed in (b) accordingly. (c) Bulging amplitudes $a_0-a_L$ varying with the ratio of  the surface tension to its critical value $\bar{\gamma}/\bar{\gamma}_\text{cr}$. The scattered hollow dots in (a) and (c) are the 3d FE results  for $\bar{\mu}_\text{s}=0.2$.   }\label{fig:solution42}
	\end{figure}
	
For $\hat{\lambda}=1.5$, the critical values of $\bar{\gamma}$ for bulging initiation are $\bar{\gamma}_\text{cr}=5.85$, $8.16$, $10.48$ accordingly. It implies that the cylinder under surface stress with higher $\bar{\mu}_\text{s}$ is less susceptible to instability. The dependence of $\bar{\gamma}$ on the stretch $\lambda_0$, the bulging evolution patterns and the necking amplitude varying with $\bar{\gamma}/\bar{\gamma}_\text{cr}$ are shown in Fig. \ref{fig:solution42}.  It sees that at the same bulging extent ($\bar{\gamma}/\bar{\gamma}_\text{cr}$), the amplitude is greater at higher  $\bar{\mu}_\text{s}$. This indicates that tough increasing the surface elasticity can raise the bulging condition, it will make the bulging more serious if the instability is triggered. 
	
Another important feature is that in the near-critical bifurcation region,  the responses of bulging amplitude $a_0-a_L$ to $\bar{\gamma}/\bar{\gamma}_\text{cr}$ first slightly increase, soon decrease, and finally increase, suggesting strain softening occurs in this scenario. From the bulging patterns shown in Fig. \ref{fig:solution42}(b), it can be observed that due to the strain softening, the bulging degree at State $3$ is worse than that at State $1$ although the surface tension is the same at the two states. This strain softening phenomenon corresponds to a subcritical bifurcation type of the instability, which was theoretically discovered by Lestringant and Audoly \cite{LA2020a} and Fu et al. \cite{Fu2021} for liquid-like cylinders. 
	
We point out that in \cite{HB2014}, the elasto-capillary instability with constant surface stress is simulated by using FEM. The softening response was not captured by their numerical scheme. Nevertheless, our FE scheme of the 3d model based on the Rayleigh-Ritz method can trace the strain softening effectively, agreeing favorably with the 1d gradient model.  Here, our results provide numerical and analytical evidence for  the subcritical bifurcation and prove that it can also occur when the surface stress is strain-dependent. Higher $\bar{\mu}_\text{s}$ will induce more prominent strain softening.
	
In addition, we find that the occurrence of the strain softening depends significantly on the length-radius ratio (see the supplementary material). The strain softening will occur in sufficiently slender cylinders (in the current configuration).  With $\hat{\lambda}=1.5$ and $L/A=40$ (i.e. $l/A=60$), the strain softening is inevitable via adjusting the surface-elasticity number. Nonetheless, it won't appear for $l/A=30$.
	
\subsubsection{Effect of surface compressibility}

\begin{figure}[ht!]
		\centering
		\includegraphics[width=0.8\linewidth]{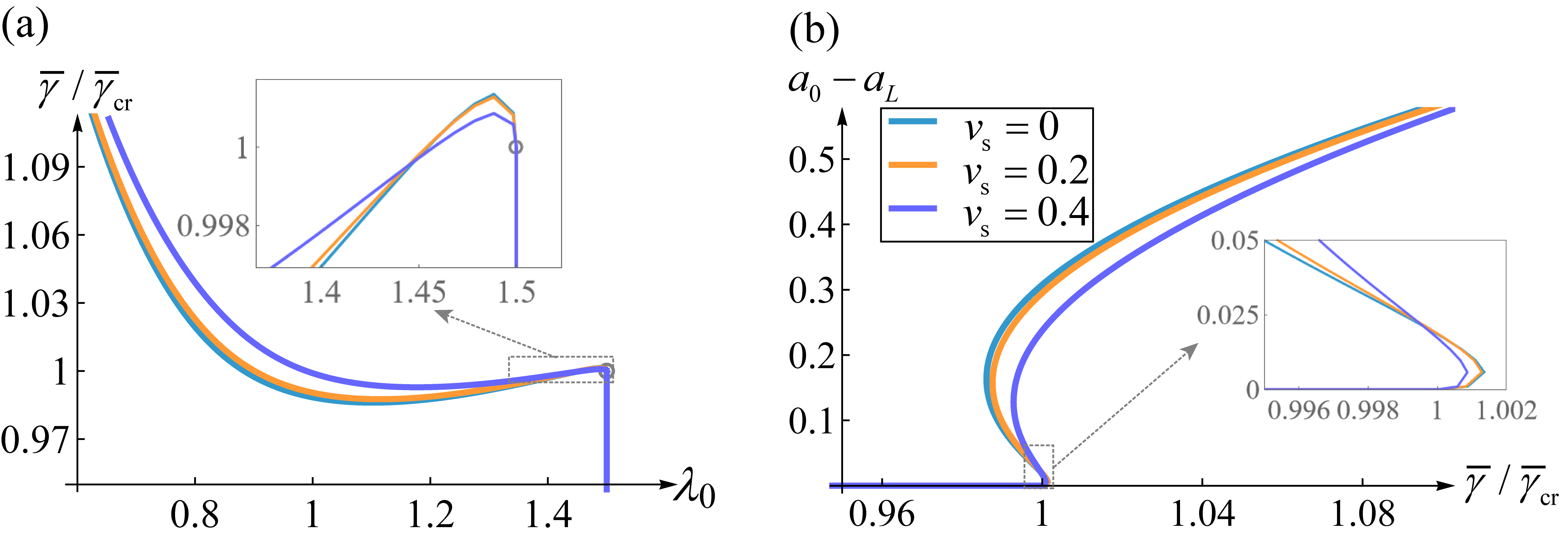}
		\caption{Bulging responses for the cases of fixed $\hat{\lambda}=1.5$ and $\bar{\mu}_\text{s}=0.2$, $\nu_\text{s}=0$, $0.4$, $0.8$ respectively, and $L/A=40$. (a) Dependence of the elasto-capillary number $\bar{\gamma}$ on the stretch $\lambda_0=\lambda(0)$ at the bulging starting point.  (b) Bulging amplitudes $a_0-a_L$ varying with the ratio of  the surface tension to its critical value $\bar{\gamma}/\bar{\gamma}_\text{cr}$.}\label{fig:solution43}
\end{figure}

Next, we set $\bar{\mu}_\text{s}=0.2$, $\nu_\text{s}=0$, $0.4$, $0.8$ respectively and $L/A=40$. For these cases, $\lambda_{\text{min}}=1.2101,~1.2189,~1.2631$ accordingly. Thus, we again take $\hat{\lambda}=1.5$ to look into the effect of surface compressibility on the bulging responses, as shown in Fig. \ref{fig:solution43}.

The critical elasto-capillary numbers at the bulging initiation for the three surface Poisson's ratios are $\bar{\gamma}_\text{cr}=8.0833$, $8.3011$, $9.3897$  accordingly. This implies that a cylinder with a more compressible surface (i.e. smaller $\nu_\text{s}$) is more prone to bulge. Yet, with smaller $\nu_\text{s}$, the same $\bar{\gamma}/\bar{\gamma}_\text{cr}$ will generate larger amplitude. This effect is almost opposite for the necking situation, except at the near-critical regime (see the supplementary material). Strain softening also occurs for each the setting of $\nu_\text{s}$.
	
Another phenomenon is that when $\nu_\text{s}$ is small (e.g., $0\leq \nu_\text{s}\leq 0.4$), both the necking and bulging amplitudes for varying $\bar{\gamma}/\bar{\gamma}_\text{cr}$ almost overlap each other in the full bifurcation regime. This indicates that if the surface possesses sufficiently high  compressibility, the post-bifurcation behavior can hardly be tuned via adjusting $\nu_\text{s}$.

\subsection{Potential application: calibration of surface parameters}

Through the above discussion, we can summarize the effects of the surface parameters in the loading scenarios as follows:
\begin{enumerate}[label=(\roman*)]
		\item With fixed surface parameters and increasing axial force, high surface-elasticity number and surface incompressibility will hinder necking initiation, but advance necking growth once the necking is triggered. The effect of surface tension is opposite.
		\item With fixed axial force and increasing elasto-capillary number, high surface-elasticity number and surface incompressibility can restrain both bulging initiation and growth, contrary to the effect of the fixed axial force.
		\item In the third loading scenario, the value of the fixed stretch determines the bifurcation type (i.e., necking or bulging). For bulging, surface elasticity boosts amplitude and strain softening; surface incompressibility acts oppositely and nonlinearly. For necking, surface elasticity elevates bifurcation condition but curbs strain softening and minimally alters amplitude; surface incompressibility delays bifurcation and amplifies amplitude and strain softening. 
\end{enumerate}
	
These findings are beneficial to recording the PR instability in solid-like materials experimentally. The instability will not occur when  $\bar{\mu}_\text{s}=\mu_\text{s}/{\mu A}$ is exceedingly large. On the contrary, it has been well known that the PR instability can be triggered at a relatively large elasto-capillary number $\bar{\gamma}=\gamma/{\mu A}$. This suggests that to witness the PR instability in a bulk-surface system with a specific surface material (characterized by a set of ($\gamma$,  $\mu_\text{s}$, $\nu_\text{s}$)), one should consider a cylinder with carefully balancing the bulk shear modulus and the radius. What's more, surface elasticity may play opposite effects on the instability initiation and growth, so do surface tension and  surface incompressibility. To record evident fluctuations, careful selection of the surface parameter values is extremely crucial. The 1d model can provide precise guidance for the selection of surface and bulk material parameters and geometrical size. 
	\begin{figure}[ht!]
		\centering
		\includegraphics[width=0.6\linewidth]{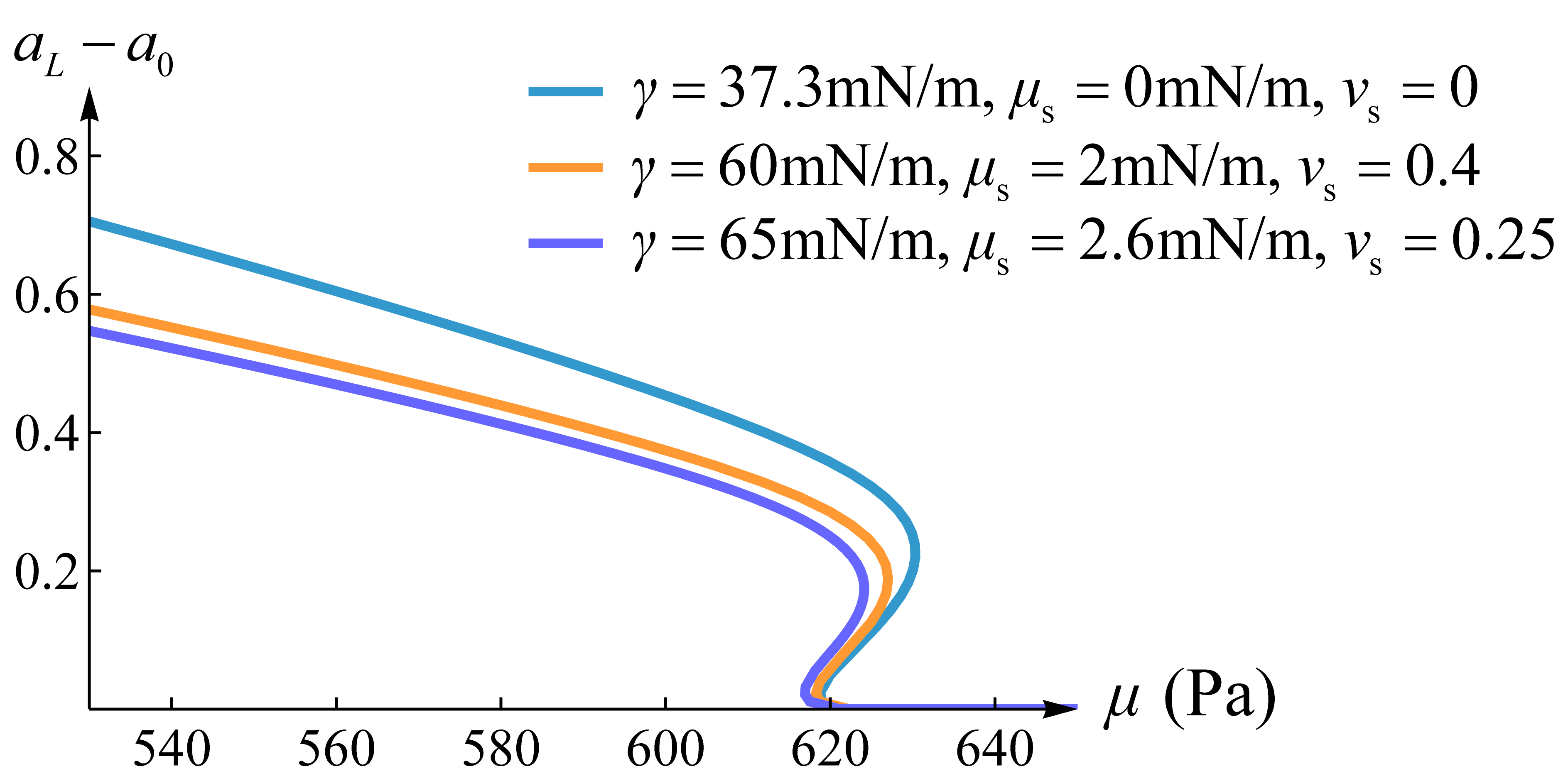}
		\caption{Illustration of calibrating surface parameters via the amplitude varying with the bulk shear modulus $\mu$. The three sets of the surface parameters $(\gamma,\mu_\text{s},\nu_\text{s})$ yield the same critical bulk shear modulus but evidently different amplitude responses. }\label{fig:mu}
	\end{figure}
	
We hereinafter elucidate a potential application of the 1d model in calibrating surface parameters. As is known, it is not easy to record the critical bifurcation state in experiments. In contrast, measuring the final amplitude of fluctuations is much simpler.  Utilizing the current 1d model, we can readily compare the amplitude response with experimental data across the entire bifurcation regime. This enables the extraction of surface parameters from experimental data.  
	
Since it may be more convenient to vary the bulk shear modulus $\mu$ to tune the PR instability in experiments (as in \cite{Mora2010}), we provide an illustration to show that the surface parameters can be identified by using the relationship between amplitude and $\mu$. For a cylinder with fixed $A=10\mu$m, $L/A=40$,  $\hat{\lambda}=1$, by adjusting the set of surface parameters (i.e. the surface tension $\gamma$, the surface shear modulus $\mu_\text{s}$ and the surface Poisson's ratio $\nu_\text{s}$), we can obtain different amplitude  responses varying with $\mu$.  As depicted in Fig. \ref{fig:mu}, we select three sets of the surface parameters. The resulting critical shear modulus is nearly identical, yet the amplitude responses exhibit different profiles. Once the experimental data of a solid-like material is given, the three surface parameters can be easily identified  through our 1d model. It is worthy pointing out that we find that the model with a constant surface stress (with $\mu_\text{s}=0$, $\nu_\text{s}=0$) is more suitable to fit the experimental data given in \cite{Mora2010}. It indicates that the surface stress in the materials used in \cite{Mora2010} is nearly strain-independent. While this study establishes a theoretical foundation for strain-dependent surface effects on PR instability, experimental validation remains an open challenge. Such experiments are beyond the scope of the current theoretical study but represent a critical future direction.
	
Building on the above insights, our demonstration of surface elasticity's fundamental role in governing PR instability in elastic solids may inspire new design principles for functional materials and structures in fields related with surface effects such as wearable sensors \cite{Li2024a}, bioengineered scaffolds \cite{G2024a}, cellular mechanical properties \cite{B2018,Camp2014}, optimization of drug delivery systems \cite{Ang2020}, micro-soft robots \cite{Roncer2024,Gosht2025} etc. For instance, as demonstrated by Goshtasbi et al. \cite{Gosht2025}, the PR instability can be utilized to develop 3d printing technique for bonding soft and rigid materials in soft robotics. The PR instability occurs during underextrusion in fused deposition modeling 3d printing, making the extruded material to form alternating necks and bulges. The authors leverage this to create a porous fibrous structure. This microstructure is key to achieving mechanical interlocking between rigid materials and soft silicones but requires precise control. Our study on surface effects, especially how surface elasticity regulates PR instability, may help optimize the 3d printing parameters (such as flow rate and printing speed) for better control of the porous structure's morphology. This could further improve the reliability of the bonding interface, which is important for the fabrication of soft robotic components like joints and grippers that require stable soft-rigid connections. We therefore hope that our theoretical findings might offer a foundation for exploring potential practical applications in these and related domains.

\section{Conclusion}\label{sec:con}

In this paper, we derived a 1d gradient model to characterize the PR instabilities in a hyperelastic cylinder with strain-dependent  surface effect. The 1d model was reduced from 3d finite-strain elasticity formulation with a generic strain energy function and surface energy function. It has been checked that the derived model can recover the one derived by Lestringant and Audoly \cite{LA2020a}, in which the surface stress is constant, independent of the surface deformation. 
		
As an application, a surface energy function with surface parameters measuring the surface tension, surface elasticity and surface compressibility was considered for the surface superelasticity, whilst the incompressible neo-Hookean model was adopted for the bulk material for computation brevity. Via the 1d reduced model, we can easily recover the bifurcation condition that is directly derived from the 3d formulation by incremental stability analysis \citep{BD2023}. Notably, our approach diverges from the prior linear analyses \citep{BD2023} by resolving the fully nonlinear post-bifurcation regime through both analytical (1d) and numerical (3d FE) treatments. More speciffically, we revealed the fully nonlinear post-bifurcation behaviors of strain-dependent surface-stress-induced/affected necking/bulging instability through simple numerical treatment of the 1d model (e.g. the finite difference method). The results of the 1d model were validated  by means of the FE scheme proposed for the 3d model. One significant feature of the FE scheme is that no imperfections needed to be introduced in the numerical simulations of 3d deformations. Both the 1d model and the FE scheme can capture the softening response occurring near the bifurcation point.
		
Three loading scenarios have been investigated. The effects of surface elasticity, surface compressibility, surface tension, axial force and length-radius ratio on the localized necking or bulging in the fully nonlinear regime were elaborated on. We found that the effects of the surface parameters on the instability are more complicated than those as concluded in \cite{BD2023}. The parameters exert different influences in different loading situations, play distinct roles in the initiation and growth of necking/bulging, and exhibit significant non-linear effects. The explorations indicate that to fine-tune the elasto-capillary behaviors in solid structures, the surface parameters and geometrical dimensions need to be carefully designed for specific application scenarios. Our 1d model and FE scheme can provide precise guidance to record the PR instability and can be applied to calibrate the surface parameters for solid-like materials.  
		
Finally, we note that in a very recent paper, Magni and Riccobelli \cite{magni2025elastic}  studied the PR instability in soft cylinders subjected to strain-dependent surface tension. Their focus is to show that the finite wavelength instability observed in experiment arises from the surface pre-stretch and strain-dependent surface tension, which does not coincide with ours. The aim of our paper is to explore of the effects of surface elasticity on the PR instability and their possible applications.

A Mathematica code that produces the results presented in this paper is available on GitHub (\href{https://github.com/pingpingzhu1/necking-and-bulging-in-soft-elastic-cylinders-with-strain-dependant-surface-effects}
	{https://github.com/pingpingzhu1})
	
\section*{Acknowledgements}
	This work was supported by the National Natural Science Foundation of China (Grant Nos. 12372077, 12402068, T2293720/T2293722), and the Guangdong Basic and Applied Basic Research Foundation (Grant Nos. 2024A1515010018, 2023A1515111141). The authors thank Professor Yibin Fu at Keele University for valuable discussions.
	
\appendix
\section{Truncation of principal stretches}\label{app:trunction}

This section presents the truncation of the principal stretches for the bulk and the outer surface of the cylinder by applying the expansion \eqref{eq:asysol}. 
		
The deformation gradient truncated at  $O(\varepsilon^2)$ is calculated as
		\begin{align}
		\bm{F}=\begin{pmatrix}
		a(S)+\varepsilon^2{u^*}/{R} & 0 & 0\\
		0 & \lambda(S)+\varepsilon^2 v^*_S & \varepsilon v^*_R\\
		0 & \varepsilon R a'(S) & a(S)+\varepsilon^2 u^*_R
		\end{pmatrix}.
	\end{align}
Accordingly, the three principal stretches, which are the eigenvalues of $\sqrt{\bm{F}^\text{T}\bm{F}}$, are  given by
		\begin{align}
		\begin{split}\label{eq:lambda}
		&\lambda_1=a+\varepsilon^2\frac{u^*}{R}, \\
		&\lambda_2=\lambda+\varepsilon^2\Big( \frac{\lambda (v^{*2}_R+R^2 a'(S)^2 )+2 R a a'(S) v^*_R}{2(\lambda^2-a^2)}+v^*_S\Big),\\
		&\lambda_3=a+\varepsilon^2\Big(-\frac{a (v^{*2}_R+R^2 a'(S)^2 )+2 R \lambda a'(S) v^*_R}{2(\lambda^2-a^2)}+u^*_R\Big).
		\end{split}
		\end{align}
A similar calculation making the use of \eqref{eq:FsF} shows that the two principal stretches of the outer surface are
		\begin{align}
		\begin{split}\label{eq:lambdas}
		&\lambda^\text{s}_1=a+\varepsilon^2\frac{u^*(S,A)}{A},\\
		&\lambda^\text{s}_2=\lambda+\varepsilon^2\Big(\frac{A^2 a'(S)^2}{2\lambda}+v^*_S(S,A)\Big).
		\end{split}
        \end{align}
	
\section{Simplification of the next-order energy $\mathcal{E}_2$}\label{app:simp}
	
This section provides the details of simplifying $\mathcal{E}_2$ to the form \eqref{eq:E2d}.
	
Upon integration by parts, one has
	\begin{align}\label{eq:uZA}
	\begin{split}
	\int_{-L}^L \varGamma_2 Av_Z(Z,A)\,dZ&=\varGamma_2 Av(Z,A)|_{Z=-L}^{Z=L}-\int_{-L}^L (\varGamma_{12}a_{\lambda}+\varGamma_{22})\lambda'(Z)Av(Z,A)\,dZ,\\
	&=\varGamma_2 Av(Z,A)|_{Z=-L}^{Z=L}-\int_{-L}^L \frac{\varGamma_{12}a_{\lambda}+\varGamma_{22}}{A}\lambda'(Z)\int_0^R(R^2v_{R}+2v R)\,dR,
	\end{split}
	\end{align}
where $a_\lambda={da}/{d\lambda}$,  $\varGamma_{12}={\partial^2 \varGamma}/{\partial\lambda^\text{s}_1\partial\lambda^\text{s}_2}$, $\varGamma_{22}={\partial^2 \varGamma}/{{\partial(\lambda^{s}_2)}^2}$, and the derivatives  are evaluated at $(a(Z),\lambda(Z))$. Note that  the following equality has been used implicitly
	\begin{align}
	A^2v(Z,A)=\int_{0}^A (R^2v_{R}+2v R)\,dR.
	\end{align}
Inserting \eqref{eq:uZA} into \eqref{eq:EE}, we see that $\mathcal{E}_2$ is simplified into
	\begin{align}
	\begin{split}\label{eq:E2b}
	\mathcal{E}_2=&\int_{-L}^L\Big[ \int_0^A \Big( w_2 \frac{\lambda (v^{2}_R+R^2 a_\lambda^2 \lambda'(Z)^2 )+2 R a a_\lambda \lambda'(Z) v_R}{2(\lambda^2-a^2)}\\
	&+\frac{\varGamma_1}{A} \frac{a (v^{2}_R+R^2 a_\lambda^2 \lambda'(Z)^2 )+2 R \lambda a_\lambda \lambda'(Z) v_R}{2(\lambda^2-a^2)}-\frac{\varGamma_{12}a_{\lambda}+\varGamma_{22}}{A}\lambda'(Z) Rv_R\Big)R\,dR\\
	&+\varGamma_2  \frac{A^3 a_\lambda^2\lambda'(Z)^2}{2\lambda}\Big]\,dZ+\varGamma_2 A v(Z,A)|_{Z=-L}^{Z=L},
	\end{split}
	\end{align}
where we have used \eqref{eq:implicit} to eliminate $w_1$. Rearranging\eqref{eq:E2b} yields the compact form of $\mathcal{E}_2$ announced in \eqref{eq:E2d}.

\section{Expression of the 1d model for neo-Hookean material models}\label{app:neo-Hookean}

The expression of the 1d model for the material models specified by \eqref{eq:W} and \eqref{eq:Gamma} is given here.
	
On substituting \eqref{eq:W} and \eqref{eq:Gamma} into \eqref{eq:Ehom} and \eqref{eq:BB}, we obtain the homogeneous potential energy $\varPhi$ and gradient modulus $B$ of the 1d model as
	\begin{align}
	\begin{split}
	&\varPhi(\lambda)=A^2\mu\Big(
	\frac{\lambda^3+4\bar{\gamma}\lambda^{3/2}-3 \lambda +2}{4
		\lambda }+\frac{\bar{\mu}_\text{s} (2 \lambda
		^3-4 \lambda -2 \lambda  \ln \lambda +2)}{4 \lambda }\\
	&\hspace{4.8em}+\frac{\bar{\mu}_\text{s}\nu_\text{s}(\lambda^2-\lambda -\lambda  \ln \lambda)}{2(1-\nu_\text{s}) \lambda
	}-\frac{\lambda \bar{N}}{2}\Big),
	\end{split}\\
	\begin{split}
	&B(\lambda)= A^4 \mu  \Big(\frac{\lambda ^3+4\bar{\gamma}\lambda^{3/2}-1}{16 \lambda
		^6}-\frac{\bar{\mu}_\text{s}(3
		\lambda +1) }{8 \lambda^6}
	-\frac{\bar{\mu}_\text{s}^2 (2\lambda^3+\lambda+1)^2}{16 \lambda
		^6}\\
	&\hspace{3.4em}+\frac{\bar{\mu}_\text{s}\nu_\text{s}(\lambda-3-\bar{\mu}_\text{s}(\lambda+1)(2\lambda^3+\lambda+1))}{8(1-\nu_\text{s})\lambda^5}
	-\frac{\bar{\mu}_\text{s}^2\nu_\text{s}^2 (\lambda+1)^2}{16(1-\nu_\text{s})^2\lambda ^4}\Big).
	\end{split}
	\end{align}

\end{document}